\documentclass[12pt]{article}
\usepackage{amsthm,amsfonts}
\swapnumbers
\theoremstyle{plain}
\newtheorem{thm}{THEOREM}
\newtheorem{lm}[thm]{LEMMA}

\newtheorem{prop}[thm]{PROPOSITION}
\theoremstyle{definition}
\newtheorem{defi}[thm]{DEFINITION}
\theoremstyle{remark}

\newtheorem{remark}[thm]{REMARK}

\newcommand{\R}{{\mathord{\mathbb R}}}
\renewcommand{\S}{{\mathord{\mathbb S}}}
\newcommand{\C}{{\mathord{\mathbb C}}}
\newcommand{\Z}{{\mathord{\mathbb Z}}}

\newcommand{\sfrac}[2]{{\textstyle \frac{#1}{#2}}}
\newcommand{\be}{\begin{equation}}
\newcommand{\ee}{\end{equation}}
\def\wt{\widetilde}
\def\Tr{{\mathop{\rm Tr}}}
\def\set#1#2{{\left\{#1 : #2\right\}}}

\def\epsilon{{\varepsilon}}

\title{The kernel of Dirac operators on $\S^3$ and $\R^3$}
\author{L\'aszl\'o Erd\H{o}s
 \thanks{L. Erd\H{o}s was supported by the N.S.F. grant DMS-9970323}\\
School of Mathematics \\ 
GeorgiaTech \\
Atlanta GA-30332, USA \\ \\
\and
Jan Philip Solovej\thanks{J.P.\ Solovej was 
supported in parts by the EU TMR-grant FMRX-CT
  96-0001 by  MaPhySto -- Centre for Mathematical Physics and 
Stochastics, funded by a grant from The Danish National Research
Foundation and by a grant from the Danish Natural Science Research
Council} \\
Department of Mathematics\\
University of Copenhagen \\
Universitetsparken 5, \\
DK-2100 Copenhagen, Denmark\\
}
\date{January 21, 2000}
\begin{document}
\baselineskip=.28in
\maketitle

\begin{abstract}
In this paper we describe an intrinsically geometric way 
of producing magnetic fields on $\S^3$ and $\R^3$
for which the corresponding Dirac operators have a non-trivial kernel.
In many cases we are able to compute the dimension of the kernel. 
In particular we can give examples where the kernel has any 
given dimension. 
This generalizes the examples of Loss and Yau \cite{LY}. 
\end{abstract}


\medskip\noindent 
{\it Running title:} Kernel of Dirac operators

\section{Introduction}

In \cite{LY} Loss and Yau proved the existence
of a magnetic field ${\bf B}=\nabla\times{\bf A}:\R^3\to\R^3$ 
with the property that $\int_{\R^3}|{\bf B}|^2<\infty$
and such that the Dirac operator
\begin{equation}\label{eq:diracr3}
        \sigma\cdot(-i\nabla -{\bf A})
\end{equation}
has a nonvanishing kernel 
in $L^2(\R^3;\C^2)$.

The significance of this result 
was its implications to the stability of matter (electrons and nuclei)
coupled to classical electromagnetic fields. 
This model was studied in the series of papers \cite{FLL,LY,LL}.
The existence of a square integrable zero mode for the Dirac
operator corresponding to a square integrable magnetic field
implies that matter cannot be stable unless there is an upper
bound on the fine structure constant.

Loss and Yau gave a very explicit construction of a magnetic 
field ${\bf B}=\nabla\times{\bf A}$ and of a 
corresponding zero mode, i.e. a solution to
$[\sigma\cdot(-i\nabla -{\bf A})]\psi=0$. They also discussed a
general way of constructing a vector potential ${\bf A}$
for a given $\psi$ so that $\psi$ be in the kernel of
the Dirac operator (\ref{eq:diracr3}). 
However their methods
gave only one element of the kernel for each magnetic field
constructed. Moreover, the proofs are very computational
and they somewhat left the origin of these zero modes unexplored.
Further examples of zero modes were given later in \cite{E}
and \cite{AMN1}, based upon ideas from the Loss and Yau construction.

In this paper we discuss a more geometric way of constructing
Dirac operators with a non trivial kernel on $\R^3$.
%In particular, we show that one can construct a Dirac operator
%with kernel of any given dimension. The same result has been 
%shown recently in \cite{AMN2} by a different construction.
More precisely, we describe a family of magnetic fields 
on the 3-sphere $\S^3$ for which we can give a 
characterization of the spectrum and in particular 
for some of these fields we can also calculate the dimension 
of the kernel. It is a well known fact (see \cite{hitchin} and
Theorem~\ref{thm:conformaldirac} below) that the dimension
of the kernel of the Dirac operator is a conformal invariant. 
Since $\R^3$ is conformally invariant to 
the 3-sphere with a point removed we can use the construction 
on $\S^3$ to learn about the kernel of Dirac operators
on $\R^3$.

On a general Riemannian manifold one can define the Dirac operator
if one has a $Spin$ structure, a corresponding spinor bundle,
and an appropriate covariant derivative (a $Spin$ connection). 
The kernel of this (nonmagnetic) Dirac operator has been studied in \cite{B}.
If one is interested in Dirac operators with magnetic fields
one must consider instead $Spin^c$ structures,
$Spin^c$ spinor bundles and a $Spin^c$ connections. The magnetic field
is then related to the curvature of the connection
(see Definition~\ref{defi:curvature}).
%We shall introduce these structures
%and define the Dirac operator for 2 and 3-dimensional
%manifolds briefly in Sect.~\ref{sec:2}. 
On $\R^3$ these structures reduce to the well known objects. 
The $Spin^c$ spinors are simply maps from $\R^3\to\C^2$ 
and the Dirac operators are of the form (\ref{eq:diracr3}).

On 2-dimensional manifolds and in general on even dimensional
manifolds the Atiyah-Singer Index Theorem often 
gives nontrivial information on the index of the Dirac
operator. In certain cases one knows from vanishing 
theorems that the index is equal to the dimension 
of the kernel. One example of such a result is 
the Aharonov-Casher Theorem (see Theorem~\ref{thm:AC})
which holds for Dirac operators on  $\R^2$ and $\S^2$. 
Characteristic for the index theorem is of course that 
the index is expressed in terms of topological quantities, 
whereas in general the dimension of the kernel 
is not a topological invariant. 
For odd-dimensional manifolds it is not easy to 
get information about the dimension of the kernel 
from index theorems. Given a Dirac operator (with   
magnetic field) on $\S^3$ we do not know in general 
how to say anything about its kernel.

In this paper we explain how, for certain magnetic fields
on $\S^3$
one may, in a sense, separate variables and reduce the problem
to a problem on $\S^2$, where one can use the Aharonov-Casher
Theorem, and a problem on $\S^1$ which can be solved explicitly.

Our construction can be used on other
manifolds than $\S^3$.

In Sect.~\ref{sec:spinbundle} we discuss $Spin^c$ structures and
define the Dirac operator and magnetic fields. 
In Sect.~\ref{sec:conformal} we discuss how the Dirac operator 
changes under conformal transformations.
In Sect.~\ref{sec:lift} we describe how to lift 
$Spin^c$ structures from 2 to 3-dimensional manifolds. 
One can of course not always do this. It requires that there is 
a map of the type known as 
a Riemannian submersion from the 3 to the 2-dimensional manifold. 
If such a map exists we give a lower bound on the kernel 
of the 3-dimensional Dirac operator using the index Theorem
for the 2-dimensional Dirac operator (see Theorem~\ref{thm:lowerbound} in
Sect.~\ref{sec:lowerbound}).
Finally in Sect.~\ref{sec:exact} we give the more detailed results 
for $\S^3$. In this case we use the Hopf map 
as the Riemannian submersion from $\S^3$ to $\S^2$.
The Hopf map however has much stronger properties than 
just being a Riemannian submersion. These properties
allow us to separate variables for the Dirac operator on 
$\S^3$. Our main results for $\S^3$ can be found in 
Theorem~\ref{thm:exact} and the remarks following it.

In particular, we show that one can construct  Dirac operators
on $\R^3$ and $\S^3$ having kernels of any given dimension. 
Examples of Dirac operators on $\R^3$ with degenerate
kernels were recently given independently in \cite{AMN2}
for a subclass of the magnetic fields considered here.
The exact degeneracy was however not proved there.

Our results were announced in \cite{ES}.

\section{$Spin^c$ bundles}\label{sec:spinbundle}

\begin{defi}[$Spin^c$ spinor bundle] Let $M$ be a 3-dimensional
Riemannian manifold. A $Spin^c$ {\it spinor bundle} $\Psi$ over $M$ is
a 2-dimensional complex vector bundle over $M$ with inner product
and an isometry $\sigma:T^*M\to\Psi^{(2)}$, where
$$\Psi^{(2)}:=
  \set{A\in\hbox{\rm End}(\Psi)}{A=A^*,\Tr A=0}.
$$
The inner product on $\Psi^{(2)}$ is given by
$(A,B):=\sfrac{1}{2}\Tr[AB]$. 
A spinor bundle over a 2-dimensional 
%(not necessarily oriented)
Riemannian manifold 
is defined in the same way except that then $\sigma$ 
is only an injective partial isometry.
The map $\sigma$ is called the {\it Clifford multiplication} of the
spinor bundle $\Psi$.
\end{defi}
Note that if 
$A,B\in\Psi^{2}$  then  
$\{A,B\}:=AB+BA=\Tr[AB]I
=2(A,B)I$ and therefore
\begin{equation}
  \label{eq:jordan}
  \{\sigma(\alpha),\sigma(\beta)\}=2(\alpha,\beta)I,\quad\mbox{for all 
    } \alpha,\beta\in T^*M
\end{equation}
where $(\cdot,\cdot)$ in the last equation denotes the 
metric (inner product) on $T^*M$. 
We have here used the convention that the Clifford multiplication 
is Hermitian rather than anti-Hermitian, which is the more common 
in the mathematics literature. 

\begin{defi}[ {\it Spin} spinor bundle] A $Spin^c$  spinor bundle $\Psi$ 
over a 2 or 3-dimensional manifold $M$ is said to be a {\it Spin} {\it spinor
 bundle}
if there exists an antilinear bundle isometry
$
        {\cal C}:\Psi\to\Psi
$
such that $(\eta, {\cal C}\eta)=0$ and ${\cal C}^2
 \eta=-\eta$ for all $\eta\in\Psi$.
In the physics literature the map ${\cal C}$
is often referred to as {\it charge conjugation}.
An equivalent way to say that a $Spin^c$ bundle is actually
a Spin bundle is to say that the {\it determinant line bundle}
$\Psi\wedge\Psi$ is trivial. 
%The charge conjugation operator
%then plays the same role as the Hodge dual for orientable 
%Riemannian manifolds.  
\end{defi}

\begin{remark} We shall use mostly   $Spin^c$ spinor bundles in our results.
For brevity, we shall refer to them simply as {\it spinor bundles}.
{\it Spin} spinor bundles will be mentioned only in some 
additional remarks.
\end{remark}

%For a 3-dimensional non-orientable manifold $M$ one cannot use the above 
%definition. In fact, as a consequence of 
The following
proposition shows that in 3-dimensions
the spinor bundle  gives a natural 
orientation of $M$. 

\begin{prop}\label{prop:onb} Let $\Psi$ be a  spinor bundle over 
a 3-dimensional manifold $M$ with Clifford multiplication $\sigma$.
If $e^1,e^2,e^3$ is an orthonormal basis in 
$T^*_pM$ then $i\sigma(e^1)\sigma(e^2)\sigma(e^3)=\pm I$. 
\end{prop}
\begin{proof} Let $a=i\sigma(e^1)\sigma(e^2)\sigma(e^3)$. It follows 
immediately from (\ref{eq:jordan}) that $a$ is Hermitian and
commutes with  $\sigma(e^j)$, $j=1,2,3$. Hence $a$ is a real scalar. 
Again from (\ref{eq:jordan}) it is clear that $\sigma(e^j)^2=I$, 
for $j=1,2,3$ and hence $a^2=I$. 
\end{proof}

\begin{defi}[Positive orientation]
We say that $e^1,e^2,e^3$ is a positively oriented
basis if $i\sigma(e^1)\sigma(e^2)\sigma(e^3)=-I$.
\end{defi}

\begin{remark}[Spinors over non-orientable manifolds]
Proposition \ref{prop:onb} shows that a spinor bundle with $\C^2$ fibers
can exist only on orientable manifolds.
However, one may also define spinor bundles over {\it non-orientable} 
3-dimensional
manifolds but in this case it should  be a $\C^4$ bundle and the
Clifford multiplication map should be an injective partial 
isometry satisfying (\ref{eq:jordan}).
\end{remark}

\begin{prop}[Basis for $\Psi$ gives basis for $T_*M$]
  \label{prop:basis}
Let $\xi_\pm$ be a local orthonormal basis of spinor fields. 
Define the vectors $e_1,e_2$  such that 
$$
    \alpha(e_1)+i\alpha(e_2)=(\xi_-,\sigma(\alpha)\xi_+)
$$
holds for all one-forms $\alpha$. Then $e_1,e_2$ are orthonormal.

In the case when $M$ is 3-dimensional also define $e_3$ such that
$$
     \alpha(e_3)=(\xi_+,\sigma(\alpha)\xi_+)
$$
for all $\alpha$.
Then $e_1,e_2,e_3$ is a positively oriented orthonormal basis.
\end{prop}
\begin{proof} We treat the 3-dimensional case, the 2-dimensional case
is similar.
We prove instead that there is an orthonormal basis 
of forms $e^1,e^2,e^3$ for which $e_1$, $e_2$, $e_3$ is the dual basis. 
This follows if we show that 
$\sigma(e^1)$, $\sigma(e^2)$, $\sigma(e^3)$ are orthonormal in $\Psi^{(2)}$.
Of course we define $e^1$, $e^2$, $e^3$ by 
$e^j(e_i)=\delta_{ij}$, for $i,j=1,\ldots$. We then see from the definitions 
of $e_1$, $e_2$, $e_3$ that the matrices of
$\sigma(e^1)$, $\sigma(e^2$), $\sigma(e^3)$,
in the basis $\xi_\pm$ are the standard Pauli matrices
(note that $(\xi_+,\sigma(\alpha)\xi_+)=-(\xi_-,\sigma(\alpha)\xi_-)$
since $\sigma(\alpha)$ is traceless)
$$
   \pmatrix{0&1\cr 1&0},\quad\pmatrix{0&-i\cr i&0},\quad\pmatrix{1&0\cr0&-1}.
$$
The orthonormality and positivity
follows from simple matrix calculations.
\end{proof}

\begin{defi}[$Spin^c$ Connection] \label{defi:spinco}
A connection $\nabla$ on 
a spinor bundle $\Psi$ is said to be a $Spin^c$ connection 
if
for all tangent vectors $X\in T_*M$ we have
\begin{description}
\item{(i.)}
 \label{item:spinco1} $X(\xi,\eta)=(\nabla_X\xi,\eta)+(\xi,\nabla_X\eta)$
for all sections $\xi,\eta$ in $\Psi$. 
\item{(ii.)}
 \label{item:spinco2} $\big [ \nabla_X, \sigma(\alpha)\big] 
=\sigma(\nabla_X\alpha)$
for all one-forms $\alpha$ on $M$.
Here $\nabla_X\alpha$ refers to the Levi-Civita connection acting
on one-forms.
% and $\nabla_X\sigma(\alpha)$ is the connection
%lifted to the Clifford multiplication, i.e., given by
%$\nabla_X\left[\sigma(\alpha)\xi\right]=\left[\nabla_X\sigma(\alpha)\right]
%\xi+\sigma(\alpha)\nabla_X\xi$.
\end{description}
\end{defi}

\begin{prop}[Local expression for $\nabla_X$]\label{prop:local}
Let $\xi_\pm$ be a local orthonormal basis of spinor fields 
on a 3-dimensional manifold
and let $e_1,e_2,e_3$ be the orthonormal basis defined in 
Prop.~\ref{prop:basis}.
Then 
for all vectors $X$
\begin{eqnarray*}
  \lefteqn{\pmatrix{(\xi_+,\nabla_X\xi_+)&(\xi_+,\nabla_X\xi_-)\cr & \cr
   (\xi_-,\nabla_X\xi_+)&(\xi_-,\nabla_X\xi_-)}}&&\\
 && \\ && \\
 &=&i\sfrac{1}{2}\pmatrix{(e_1,\nabla_Xe_2)&
   -(e_3,\nabla_Xe_2)-i(e_3,\nabla_Xe_1)
   \cr &\cr
   -(e_3,\nabla_Xe_2)+i(e_3,\nabla_Xe_1)&-(e_1,\nabla_Xe_2)}
 -i\alpha(X)I,
\end{eqnarray*}
where $\alpha$ is the real local one-form given by
$$
     \alpha(X)=i\sfrac{1}{2}(\xi_+,\nabla_X\xi_+)+
     i\sfrac{1}{2}(\xi_-,\nabla_X\xi_-).
$$ 

The same formulas are true in the 2-dimensional case if we simply 
replace $e_3$ by $0$ everywhere.
\end{prop}
\begin{proof} Let $e^1,e^2,e^3$ be the dual basis to $e_1,e_2,e_3$.
Using that $X(\omega(e_j))=\nabla_X\omega(e_j)+\omega(\nabla_Xe_j)$ 
for any one-form $\omega$
we find, with $\omega=e^j$, that  
\begin{eqnarray*}
    \lefteqn{(e_j,\nabla_Xe_1)+i(e_j,\nabla_Xe_2)
    =e^j(\nabla_Xe_1)+ie^j(\nabla_Xe_2)}&&\\
    &=&X(\xi_-,\sigma(e^j)\xi_+)-(\xi_-,\sigma(\nabla_Xe^j)\xi_+)\\
    &=&(\nabla_X\xi_-,\sigma(e^j)\xi_+)
    +(\xi_-,\sigma(e^j)\nabla_X\xi_+).
\end{eqnarray*}
Since $\sigma(e^3)\xi_\pm=\pm\xi_\pm$ (see the 
proof of Prop.~\ref{prop:basis})
we find 
$$
    (e_3,\nabla_Xe_1)+i(e_3,\nabla_Xe_2)=(\nabla_X\xi_-,\xi_+)
    -(\xi_-,\nabla_X\xi_+)=-2(\xi_-,\nabla_X\xi_+)
$$
where we have also used that $0=X(\xi_-,\xi_+)=(\nabla_X\xi_-,\xi_+)
+(\xi_-,\nabla_X\xi_+)$. This also gives 
$$
    (\xi_+,\nabla_X\xi_-)=-\overline{(\xi_-,\nabla_X\xi_+)}
    =\sfrac{1}{2}(e_3,\nabla_Xe_1)-i\sfrac{1}{2}(e_3,\nabla_Xe_2).
$$
Using $\sigma(e^1)\xi_\pm=\xi_\mp$ we obtain
\begin{eqnarray*}
   (\xi_+,\nabla_X\xi_+)-(\xi_-,\nabla_X\xi_-)&=&
   (\nabla_X\xi_-,\xi_-)+(\xi_+,\nabla_X\xi_+)\\
   &=&(e_1,\nabla_Xe_1)+i(e_1,\nabla_Xe_2)=i(e_1,\nabla_Xe_2)
\end{eqnarray*}
where we have used that $(e_1,\nabla_Xe_1)=\sfrac{1}{2}X(e_1,e_1)=0$.
It only remains to show that $\alpha$ is real. 
This  follows from $0=X(\xi_\pm,\xi_\pm)=
2\mbox{Re}(\xi_\pm,\nabla_X\xi_\pm)$.

In the 2-dimensional case we conclude from $(\nabla_Xe^j,e^j)=0$
that $\nabla_Xe^1=(\nabla_Xe^1,e^2)e^2$ and
$\nabla_Xe^2=(\nabla_Xe^2,e^1)e^1$.
Since $\sigma(e^1)\sigma(e^2)\xi_+=i\xi_+$ we have
\begin{eqnarray*}
  \sigma(e^1)\sigma(e^2)\nabla_X\xi_+
  &=&\nabla_X(\sigma(e^1)\sigma(e^2)\xi_+)
  -\sigma(\nabla_Xe^1)\sigma(e^2)\xi_+-\sigma(e^1)\sigma(\nabla_Xe^2)
  \xi_+\\ 
  &=& i\nabla_X\xi_+ -
  \left((\nabla_Xe^1,e^2)+(\nabla_Xe^2,e^1)\right)\xi_+=
  i\nabla_X\xi_+,
\end{eqnarray*}
where we have used that 
$(\nabla_Xe^1,e^2)+(\nabla_Xe^2,e^1)=X(e^1,e^2)=0$.
Therefore $\nabla_X\xi_+$ is an eigenvector with eigenvalue
$i$ of the antihermitean operator $\sigma(e^1)\sigma(e^2)$.
Since $\sigma(e^1)\sigma(e^2)\xi_-=-i\xi_-$ 
we conclude that $(\xi_-,\nabla_X\xi_+)=0$ in the 2-dimensional
case. The other statements in the 2-dimensional case follow 
as for the 3-dimensional case.
\end{proof}

\begin{defi}[Curvature tensor and magnetic form on $\Psi$]
As usual the curvature tensor is defined by 
$$ 
{\cal R}_\Psi(X,Y)\xi=\nabla_X\nabla_Y\xi-\nabla_Y\nabla_X\xi-
\nabla_{[X,Y]}\xi,
$$
where $\xi$ is a spinor section and $X,Y$ are vector fields on $M$.
The {\it Magnetic 2-form} $\beta$ on $M$ is defined via
the trace of ${\cal R}_\Psi$ as
\be
  \beta(X,Y)=i\sfrac{1}{2}\Tr[{\cal R}_\Psi(X,Y)].
\label{defi:curvature}
\ee
\end{defi}

\begin{prop}\label{prop:magnfield}
 The operator $i{\cal R}_\Psi(X,Y)$ is Hermitian
on $\Psi$ and $\beta$ is a real closed two-form.
Locally $\beta = d\alpha$, where $\alpha$ is the one-form
defined in Proposition \ref{prop:local}.
\end{prop}
\begin{proof} In fact it follows from the definition
of a $Spin^c$ connection that 
$$
   X\left(Y\left((\xi,\eta)\right)\right)=
   (\nabla_X\nabla_Y\xi,\eta)+(\xi,\nabla_X\nabla_Y\eta)
   +(\nabla_X\xi,\nabla_Y\eta)+(\nabla_Y\xi,\nabla_X\eta).
$$
Hence 
\begin{eqnarray*}
    \lefteqn{(\nabla_X\nabla_Y\xi-\nabla_Y\nabla_X\xi,\eta)
    +(\xi,\nabla_X\nabla_Y\eta-\nabla_Y\nabla_X\eta)}&&\\
    &=&X\left(Y\left((\xi,\eta)\right)\right)
    -Y\left(X\left((\xi,\eta)\right)\right)\\
    &=&[X,Y]\left((\xi,\eta)\right)
    = (\nabla_{[X,Y]}\xi,\eta)+(\xi,\nabla_{[X,Y]}\eta)
\end{eqnarray*}   
and thus $({\cal R}_\Psi(X,Y)\xi,\eta)+(\xi,{\cal
  R}_\Psi(X,Y)\eta)=0$.
%\begin{prop} The magnetic 2-form $\beta$ is a real closed 
%2-form on $M$. \end{prop}

That $\beta$ is real is a consequence of 
$i{\cal R}_\Psi$ being Hermitian.
That it is a 2-form, i.e, that it is antisymmetric is immediate from the 
definition. That $\beta = d\alpha$ will be shown in the proof of
Theorem \ref{thm:spincurv}, hence it immediately follows that $\beta$
is closed.
\end{proof}

%We now prove that it is closed. 
%Let $\xi_\pm$ be a local  orthonormal basis of spinor sections.
%Let $\alpha$ be the corresponding one-form defined in 
%Prop.~\ref{prop:local}. We shall show that $d\alpha=\beta$ and hence
%that $\beta$ is closed. Note that whereas $\beta$ is globally 
%defined $\alpha$ is only defined locally, hence $\beta$ 
%is not necessarily exact. 
%We calculate
%\begin{eqnarray*}
%     2d\alpha(X,Y)&=&2X(\alpha(Y))-2Y(\alpha(X))-2\alpha([X,Y])\\
%     &=&X(\xi_+,i\nabla_Y\xi_+)+X(\xi_-,i\nabla_Y\xi_-)
%     -Y(\xi_+,i\nabla_X\xi_+)-Y(\xi_-,i\nabla_X\xi_-)
%     \\
%     &&{}-(\xi_+,i\nabla_{[X,Y]}\xi_+)-(\xi_-,i\nabla_{[X,Y]}\xi_-)\\
%     &=&2\beta(X,Y)+(\nabla_X\xi_+,i\nabla_Y\xi_+)
%     -(\nabla_Y\xi_+,i\nabla_X\xi_+)\\
%     &&{}+(\nabla_X\xi_-,i\nabla_Y\xi_-)
%     -(\nabla_Y\xi_-,i\nabla_X\xi_-).
%\end{eqnarray*}
%Note however that since $(\nabla_X\xi_-,\xi_+)=-(\xi_-,\nabla_X\xi_+)$ 
%we have
%\begin{eqnarray*}
%    \lefteqn{(\nabla_X\xi_+,\nabla_Y\xi_+)+(\nabla_X\xi_-,\nabla_Y\xi_-)}&&\\
%    &=&(\nabla_X\xi_+,\xi_+)(\xi_+,\nabla_Y\xi_+)
%    +(\nabla_X\xi_+,\xi_-)(\xi_-,\nabla_Y\xi_+)\\
%    &&{}+
%    (\nabla_X\xi_-,\xi_+)(\xi_+,\nabla_Y\xi_-)
%    +(\nabla_X\xi_-,\xi_-)(\xi_-,\nabla_Y\xi_-)\\
%    &=&(\nabla_X\xi_+,\xi_+)(\xi_+,\nabla_Y\xi_+)
%    +(\xi_+,\nabla_X\xi_-)(\nabla_Y\xi_-,\xi_+)\\
%    &&{}+
%    (\nabla_X\xi_-,\xi_+)(\xi_+,\nabla_Y\xi_-)
%    +(\nabla_X\xi_-,\xi_-)(\xi_-,\nabla_Y\xi_-),
%\end{eqnarray*}
%and this is real since 
%$(\nabla_X\xi_+,\xi_+)$ is purely imaginary. 
%We therefore conclude that $d\alpha=\beta$.
%\end{proof}

\begin{thm}[Spinor curvature in terms of Riem.\ curvature\\ ]
\label{thm:spincurv}
Let ${\cal R}$ denote the Riemann curvature tensor of $M$
and let $\{e_i\}$ be any orthonormal basis of vector fields
and $\{e^i\}$ the dual basis of one-forms.
For all vectors $X$ and $Y$ we then have the identity 
$$
        {\cal R}_\Psi(X,Y)=\sfrac{1}{4}\sum_{ij}
        (e_i,{\cal R}(X,Y)e_j)\sigma(e^i)\sigma(e^j)-i\beta(X,Y)I
$$
\end{thm}
\begin{proof}
It is enough to check this identity at a point $p\in M$.
It is also clear that the identity is independent 
of the choice of orthonormal basis of vectors.
We may choose an orthonormal spinor basis $\xi_\pm$ 
such that $\nabla_X\xi_\pm=0$ at $p$ for all vectors
$X$. The corresponding vectors $e_i$ defined in Prop.~\ref{prop:basis}
gives a local geodesic basis at $p$, i.e., $\nabla_{e_j}e_i=0$.
Moreover, the one-form $\alpha$ from Prop.~\ref{prop:local}
vanishes at the point $p$.
Using the local expression for the $Spin^c$ connection given in 
Prop.~\ref{prop:local} we find
\begin{eqnarray*}
        {\cal R}_\Psi(X,Y)\xi_\pm&=&\sfrac{i}{2}\left(\pm(e_1,{\cal R}(X,Y)e_2)\xi_\pm
        -\left((e_3,{\cal R}(X,Y)e_2)\mp i(e_3,{\cal R}(X,Y)e_1)\right)
        \xi_\mp\right)\\
        &&-i\left(\nabla_X\alpha(Y)-\nabla_Y\alpha(X)\right)\xi_\pm
\end{eqnarray*}
That this agrees with the expression stated in the theorem 
follows easily from the expressions in  
Prop.~\ref{prop:basis}. We see immediately  that $\beta=d\alpha$.
If we change the basis $\xi_\pm$, the one-form $\alpha$ changes
by the addition of an exact one-form, hence $d\alpha$ is unchanged.
\end{proof}

\medskip

We end our  discussion on $Spin^c$ structures by showing 
how these results are modified for $Spin$ spinor bundles.

\begin{defi}[$Spin$ Connection] A $Spin^c$ connection $\nabla$
on a spinor bundle which is also a $Spin$ spinor bundle is said to be 
a $Spin$ connection if $\nabla$ commutes with the 
charge conjugation operator ${\cal C}$.
\end{defi}

\begin{prop}[Uniqueness of $Spin$ connections]\label{prop:unique}
If $\nabla'$ and $\nabla''$ are two $Spin^c$ connections 
on the same spinor bundle
then there is a (real) one-form $\omega$ 
such that $(\nabla'_X-\nabla''_X)\xi=i\omega(X)\xi$, 
for all vector fields $X$ and all spinor fields $\xi$.
In particular, if $\nabla'$ and $\nabla''$ are two 
$Spin$ connections on the same $Spin$ spinor bundle then $\nabla'=\nabla''$.
\end{prop}
\begin{proof} It follows immediately from the definition
of $Spin^c$ connections that 
$(\nabla'_X-\nabla''_X)$ commutes with multiplication by 
(complex) functions and with Clifford multiplication of one-forms. 
Hence $(\nabla'_X-\nabla''_X)$ is multiplication by a (complex) scalar.
{F}rom (\ref{item:spinco1}) in the Definition~\ref{defi:spinco}
of $Spin^c$ connections it follows that it is a purely imaginary scalar.
If $\nabla'$ and $\nabla''$ are $Spin$ connections then 
then multiplication by the imaginary scalar $(\nabla'_X-\nabla''_X)$
commutes with the antilinear map ${\cal C}$ hence the scalar is zero.
 \end{proof}

\begin{prop}[{\it Spin} connections are non-magnetic]
If $\nabla$ is a $Spin$ connection then the 
corresponding magnetic 2-form vanishes. 
\end{prop}
\begin{proof}
Indeed, if $\xi$ is a unit spinor we have
\begin{eqnarray*}
\beta(X,Y)&=&i\sfrac{1}{2}\Big(\left(\xi,{\cal R}_\Psi(X,Y)\xi\right)+
\left({\cal C}\xi,{\cal R}_\Psi(X,Y){\cal C}\xi\right)\Big)\\
&=&i\sfrac{1}{2}\Big(\left(\xi,{\cal R}_\Psi(X,Y)\xi\right)+
\left({\cal R}_\Psi(X,Y)\xi,\xi\right)\Big)=0 .
\end{eqnarray*}
since ${\cal C}$ is antilinear, it commutes with
${\cal R}_\Psi$ and satisfies $(\eta_1, {\cal C}\eta_2) = (\eta_2, \eta_1)$.
\end{proof}

%\begin{remark} If we use the extension of Clifford multiplication 
%to higher order forms we may write
%$$
%        \sum_{ij}
%        (e_i,{\cal R}(X,Y)e_j)\sigma(e^i)\sigma(e^j)
%        =\sigma\left(\sum_{ij}
%        (e_i,{\cal R}(X,Y)e_j)e^i\wedge e^j\right).
%$$
%\end{remark}

\section{The Dirac  and Laplace operators on Spinors}

Given a $Spin^c$ connection on a Spinor bundle $\Psi$ 
we may define the first order Dirac operator and the 
second order Laplace operator on spinor sections. 

\begin{defi}[The Dirac operator]
The Dirac operator  ${\cal D}:\Gamma(\Psi)\to\Gamma(\Psi)$
is given by 
$$
   \quad {\cal D}\xi:=-i\sum_j\sigma(e^j)\nabla_{e_j}\xi
$$
where  $\{e_j\}$ is an orthonormal basis of vectors
and $\{e^j\}$ is the dual orthonormal basis of one forms. 
Here $j$ runs from 1 to 2 or 3 depending on whether we are in the 2 
or 3 dimensional case. 
It is straightforward to see that this definition is
independent of the choice of basis $\{e_j\}$.
\end{defi}

It would maybe have been more suggestive to write 
${\cal D}=(-i)\sigma(\nabla)$.

The important observation about the Dirac operator is
that it is symmetric.
\begin{thm}The Dirac operator is symmetric, i.e., 
for any two spinor fields $\xi$ and $\eta$
$$
        \int_M(\xi,{\cal D}\eta)=\int_M({\cal D}\xi,\eta).
$$
\end{thm}
\begin{proof}
We compute the formal adjoint of ${\cal D}$,
$$
        {\cal D}^*=i\sum_j\nabla_{e_j}^*\sigma(e^j).
$$
We now use that 
$
        \nabla_{X}^*=-\nabla_X-\mbox{div}X
        =-\nabla_X-\sum_j(\nabla_{e_j}X,e_j).
$
We then obtain
$$
        {\cal D}^*
        =-i\sum_j\sigma(e^j)\nabla_{e_j}-i\sum_j\sigma(\nabla_{e_j}e^j)
        -i\sum_j\mbox{div}(e_j)\sigma(e^j).
$$
Note that 
$$
        \sum_j\nabla_{e_j}e^j=\sum_j\sum_k(\nabla_{e_j}e^j,e^k)e^k=
        -\sum_j\sum_k(\nabla_{e_j}e_k,e_j)e^k=-\sum_k\mbox{div}(e_k)e^k.
$$
and therefore the last two terms above cancel. 
\end{proof}

\begin{defi}[Laplace operator on $\Psi$]
The Laplace operator $\Delta$ on spinor fields is given by 
$$
 -\Delta\xi=
        \left(\nabla_{e_1}^*\nabla_{e_1}+\nabla_{e_2}^*\nabla_{e_2}
   +\nabla_{e_3}^*\nabla_{e_3}\right)\xi.
$$
\end{defi}

We end this section by proving the famous Lichnerowicz formula.
\begin{thm}[Lichnerowicz formula]
On spinor fields $\xi$ we have the Lichnerowicz formula
$$
        {\cal D}^2\xi=-\Delta\xi+\sfrac{1}{4}R\xi + i\sum_{i<j}\beta(e_i,e_j)
        \sigma(e^i)\sigma(e^j)\xi,
$$
where $R$ is the scalar curvature and $\{e_i\}$ is any orthonormal basis
of vector fields and $\{e^i\}$ is the corresponding dual basis
of one-forms. Using the extension of the Clifford multiplication 
to higher order forms we could simply write the last term above as
$i\sigma(\beta)$. In 3-dimensions this is identical to $-\sigma(*\beta)$, 
where the $*$ again refers to the Hodge dual.
\end{thm}
\begin{proof}
We shall prove the Lichnerowicz formula at a point $p$. 
As usual we may choose an orthonormal basis $\{e_i\}$
such that $\nabla_{e_j}e_i(p)=0$, for all $i,j$ 
and thus also $\nabla_{e_j}e^i(p)=0$, for all $i,j$,
where $\{e^i\}$ is the dual basis. 
We then have at $p$ that $\nabla_{e_i}^*=-\nabla_{e_i}$
and 
\begin{eqnarray*}
        {\cal D}^2&=&-\sum_i\nabla_{e_i}^2-
        \sum_{i<j}\sigma(e^i)\sigma(e^j)\left[\nabla_{e_i}\nabla_{e_j}
        -\nabla_{e_j}\nabla_{e_i}\right]\\
        &=&-\Delta - \sum_{i<j}\sigma(e^i)\sigma(e^j){\cal R}_{\Psi}
        (e_i,e_j)\\
        &=&-\Delta-\sum_{i<j}\sigma(e^i)\sigma(e^j)
        \left[\sfrac{1}{2}\sum_{k<l}(e_k,{\cal R}(e_i,e_j)e_l)
\sigma(e^k)\sigma(e^l)-i\beta(e_i,e_j)\right]\\
        &=&-\Delta-\sum_{i<j}\sigma(e^i)\sigma(e^j)
        \left[\sfrac{1}{2}(e_i,{\cal R}(e_i,e_j)e_j)
\sigma(e^i)\sigma(e^j)-i\beta(e_i,e_j)\right]
\end{eqnarray*}
where the last equality follows from the Bianchi
identity. 
We therefore arrive at
$$
{\cal D}^2=-\Delta+\sfrac{1}{4}\sum_{ij}(e_i,{\cal R}(e_i,e_j)e_j)
+i\sum_{i<j}\beta(e_i,e_j)\sigma(e^i)\sigma(e^j).
$$
\end{proof}

\begin{remark}
We may of course extend $-\Delta$ and ${\cal D}$ defined in
the sense of distributions to all $L^2$-sections $L^2(\Psi)$. Then $-\Delta$
and ${\cal D}$ are self-adjoint on the maximal domains
$\{ \psi\in L^2(\Psi)\; : \; \Delta\psi \in L^2(\Psi)\}$
and $\{ \psi\in L^2(\Psi)\; : \; {\cal D}\psi \in L^2(\Psi)\}$,
respectively.
\end{remark}

\section{Conformal transformations}\label{sec:conformal}

We now consider how spin structures change under conformal 
transformations (see also \cite{hitchin}). Let $g$ be the original metric 
on $M$ and let $g_\Omega=\Omega^2g$ be a conformal metric. 
Here $\Omega:M\to \R$ is a smooth non-vanishing function. 

If $\Psi$ is a spinor bundle over $M$ and $\sigma$ is the
corresponding Clifford map with respect to the metric $g$ then 
$\sigma_\Omega=\Omega^{-1}\sigma$ is Clifford map with respect
to the metric $g_\Omega$.

We are interested in how the connections change. 

\begin{prop}[Conformal change of Levi-Civita connection] 
\label{prop:conflc}
Let $\nabla$ be the Levi-Civita connection corresponding to the metric $g$
and $\nabla^{(\Omega)}$ be the Levi-Civita connection
corresponding to the metric $g_\Omega$. Then 
for all one forms $\omega$ and all vectors $X$, $Y$ we have 
\begin{equation}\label{eq:conf1}
     \nabla^{(\Omega)}_X\omega=
     \nabla_X\omega-\Omega^{-1}X(\Omega)\omega+
                                        \Omega^{-1}(\omega,d\Omega)X^*
     -\Omega^{-1}\omega(X)d\Omega
\end{equation}
and
\begin{equation}\label{eq:conf2}
        \nabla^{(\Omega)}_XY=
     \nabla_XY+\Omega^{-1}X(\Omega)Y+\Omega^{-1}Y(\Omega)X
     -\Omega^{-1}(X,Y)d\Omega^*,
\end{equation}
where $(\cdot,\cdot)$ refers to the inner product on one-forms and vectors
corresponding to the metric $g$. 
Here $X^*$ is the one-form corresponding to the vector 
$X$ with respect to the
metric $g$
and likewise $d\Omega^*$ is the vector corresponding to the 
one-form $d\Omega$, i.e, 
$X^*(Y)=(X,Y)$ and $(d\Omega^*,Y)=d\Omega(Y)=Y(\Omega)$ for all vectors $Y$.
\end{prop}
\begin{proof}
It is enough to prove (\ref{eq:conf2}) since (\ref{eq:conf1})
follows from the identity $\Omega^{-2}(\nabla^{(\Omega)}_X\omega)^*
=\nabla^{(\Omega)}_X(\Omega^{-2} \omega^*)$. 
Here the $\Omega$ factors are due to the fact that  $*$ is 
the duality in the $g$ metric and not the $g_\Omega$ metric.
Since $\nabla^{(\Omega)}$ clearly has all the properties of a connection
we only have to check that
\begin{description}
\item{(i)} it is torsion free, i.e.,
$$
        \nabla^{(\Omega)}_X Y- \nabla^{(\Omega)}_Y X=[X,Y]
$$
for any vector fields $X,Y$ ;
\item{(ii)} it is compatible with the metric $g_\Omega$, i.e.,
$$
        X\left[(Y,Z)_\Omega\right]=\left(\nabla^{(\Omega)}_XY,
        Z\right)_\Omega+\left(Y,\nabla^{(\Omega)}_XZ\right)_\Omega
$$
for any vector fields $X,Y,Z$.
\end{description}
Both of these follow from simple calculations.
\end{proof}

\begin{prop}[Conformal change of $Spin^c$ connection]
\label{prop:conformalspinconnection}
Let $\nabla$ be a $Spin^c$  connection
on a spinor bundle $\Psi$ on $M$ with Clifford map $\sigma$ 
corresponding to the metric $g$.
Then $\nabla^{(\Omega)}$ defined by 
\begin{eqnarray*}
     \nabla^{(\Omega)}_X&=&
     \nabla_X-\sfrac{1}{4}\Omega^{-1}\sigma(d\Omega)\sigma(X^*)
     +\sfrac{1}{4}\Omega^{-1}\sigma(X^*)\sigma(d\Omega)\\
        &=&\nabla_X+\sfrac{1}{4}\Omega^{-1}[\sigma(X^*),\sigma(d\Omega)]
\end{eqnarray*}
for any vector $X$,
is  a $Spin^c$ connection on the same spinor bundle  
with Clifford map $\sigma_\Omega=\Omega^{-1}\sigma$ corresponding
to the metric $g_\Omega$. Here again $X^*$ refers to the
one-form which is dual to the vector $X$ relative to the metric $g$.
\end{prop}
\begin{proof}
  We have to prove (\ref{item:spinco1}) and (\ref{item:spinco2})
in Definition~\ref{defi:spinco} of $Spin^c$ connections.
The first relation (\ref{item:spinco1}) follows easily from the fact that 
the Clifford multiplication is Hermitian.

To prove (\ref{item:spinco2}) we note that the lift of $\nabla^{(\Omega)}_X$
to the Clifford multiplication is given by the following expression 
involving a double commutator
\begin{eqnarray}
        \nabla^{(\Omega)}_X\left(\sigma_\Omega(\omega)\right)=
        \nabla_X[\Omega^{-1}\sigma(\omega)]
     +\sfrac{1}{4}\Omega^{-2}
        \left[\left[\sigma(X^*),\sigma(d\Omega)\right],\sigma(\omega)\right].
\end{eqnarray}
Using the commutator formula $[[A,B],C]=\{A,\{B,C\}\}-\{B,\{A,C\}\}$
we find
\begin{eqnarray*}
\left[\left[\sigma(X^*),\sigma(d\Omega)\right],\sigma(\omega)\right]
        &=&\{\sigma(X^*),\{\sigma(d\Omega),\sigma(\omega)\}\}
        -\{\sigma(d\Omega),\{\sigma(X^*),\sigma(\omega)\}\}\\
        &=&4(\omega,d\Omega)\sigma(X^*)-
        4\omega(X)\sigma(d\Omega).
\end{eqnarray*}
Hence 
$$
        \nabla^{(\Omega)}_X\left(\sigma_\Omega(\omega)\right)=
        \Omega^{-1}\sigma(\nabla_X\omega)-\Omega^{-2}X(\Omega)\sigma(\omega)
        +\Omega^{-2}(\omega,d\Omega)\sigma(X^*)-
        \Omega^{-2}\omega(X)\sigma(d\Omega).
$$

We see immediately from (\ref{eq:conf1}) that this agrees with
$
        \sigma_\Omega(\nabla_X^{(\Omega)}\omega)
        =\Omega^{-1}\sigma(\nabla_X^{(\Omega)}\omega).
$
\end{proof}

\begin{thm}[Conformal change of the Dirac operator]
\label{thm:conformaldirac}
Let ${\cal D}$ and ${\cal D}_\Omega$
denote the Dirac operators corresponding to
the $Spin^c$ connections $\nabla$ and $\nabla^{(\Omega)}$
as defined in Prop.~\ref{prop:conformalspinconnection}.
Then 
$$
  {\cal D}_\Omega=\left\{
        \begin{array}{ll}
        \Omega^{-3/2}{\cal D}\Omega^{1/2},&\mbox{in the 2-dimensional case}\\
        \Omega^{-2}{\cal D}\Omega,&\mbox{in the 3-dimensional case .}
        \end{array}\right. 
$$
\end{thm}
\begin{proof}
This follows from a simple calculation using 
Prop.~\ref{prop:conformalspinconnection}. Indeed, we have
$$
         {\cal D}_\Omega=-i\sum_j\sigma_\Omega(e^j_\Omega)
        \nabla^{(\Omega)}_{e_j^\Omega}.
$$
Here $e^j_\Omega=\Omega e^j$ and $e_j^\Omega=\Omega^{-1} e_j$
are respectively one-forms and vectors that are orthonormal
with respect to the metric $g_\Omega$ if
$e^j$ and $e_j$ are one-forms and vectors
orthonormal with respect to $g$.
Using $\sigma_\Omega=\Omega^{-1}\sigma$ we obtain 
from Prop.~\ref{prop:conformalspinconnection}in dimension $n=2$ or $3$
that
\begin{eqnarray*}
 {\cal D}_\Omega &=&-i\sum_j\Omega^{-1}\sigma(e^j)\nabla^{(\Omega)}_{e_j}
        =\Omega^{-1}{\cal D}-\sfrac{i}{4}\sum_j\Omega^{-2}\sigma(e^j)
        [\sigma(e^j),\sigma(d\Omega)]\\
        &=&\Omega^{-1}{\cal D}-\sum_j\sfrac{i}{2}\Omega^{-2}\left(\sigma(d\Omega)
        -\sfrac{1}{2}\sigma(e^j)\{\sigma(e^j),\sigma(d\Omega)\}\right)\\
        &=&\Omega^{-1}{\cal D}-\sfrac{i}{2}\Omega^{-2}\left(n\sigma(d\Omega)
        -\sum_j\sigma(e^j)(e^j,d\Omega)\right)\\
        &=&\Omega^{-1}{\cal D}-i\frac{n-1}{2}\Omega^{-2}\sigma(d\Omega)
        =\Omega^{-(n+1)/2}{\cal D}\Omega^{(n-1)/2}.
\end{eqnarray*}
\end{proof}

\section{Riemannian submersions}\label{sec:lift}

Having studied how spin structures and Dirac operators change under 
conformal transformations we now turn to transformations
between spaces of different dimensions.

We assume throughout this section that $M$ is a 3-dimensional and
$N$ is a 2-dimensional Riemannian manifold.
The natural type of transformations to study are Riemannian submersions
$\phi:M\to N$.
This means that  $\phi_*:T_*M\to T_*N$ is a surjective partial isometry.
For a discussion of Riemannian submersions and their relations 
to spin geometry see \cite{Gilkeyetal}.

We show that it is possible to pull back spinor bundles, $Spin^c$
connections and Dirac operators from $N$ to $M$ along a
Riemannian submersion.

We denote $vol_N$ the volume form on $N$.
Let $\nu = *\phi^*(vol_{N})$, i.e. $\nu$ is the Hodge dual
of the pull back of the volume form by the Riemannian
submersion $\phi$. In particular $\nu$ is a one-form on $M$.
Note that any one-form on $M$ is a linear combination of $\nu$
and the pull-back to $M$ of a one-form on $N$. These
properties are summarized in

\begin{prop}[The pull back of the volume form]\label{prop:pullvolform}
Let  $n$ be the
dual vectorfield to $\nu=*\phi^*(vol_{N})$.
If $f^1$, $f^2$ is a (locally defined) oriented orthonormal
basis in $T^*N$ then  
$\nu$, $\phi^*(f^1)$, $\phi^*(f^2)$ is a (locally defined) 
oriented orthonormal basis in
$T^*M$.
% and this uniquely determines $\nu$. 
Let $n$, $e_1$, $e_2$ be the orthonormal
vectors dual to the one-forms
$\nu$, $\phi^*(f^1)$, $\phi^*(f^2)$. Then 
$f_1=\phi_*(e_1)$, $f_2=\phi_*(e_2)$ is the dual basis to 
$f^1,f^2$ and we have
\begin{description}
\item{(i)} $([n,e_i],e_j)=0$ for $i,j=1,2$
\item{(ii)} $(\nabla^M_{e_i}e_j,e_k)=(\nabla^N_{f_i}f_j,f_k)$ for
  $i,j,k=1,2$, where $\nabla^M$ and $\nabla^N$ denote
  the Levi-Civita connections on $M$ and $N$ respectively.
\item{(iii)} We have the identity
\begin{eqnarray*}
  (\nu,*d\nu)&=&d\nu(e_1,e_2)=-(n,[e_1,e_2])=2(\nabla^M_n e_1,e_2)\\
   &=&2(\nabla^M_{e_1}n ,e_2).
 \end{eqnarray*}
Note that $n$ is (locally) hypersurface orthogonal if and only if this
quantity vanishes.
\end{description}
\end{prop}
\begin{proof}
The characterization of $\nu$ is straightforward from the definition. 
It is also clear that $\phi_*(n)=0$.
If $X$ is 
a vector on $M$ orthogonal to $n$ then since $\phi_*$ is an isometry 
on the subspace orthogonal to $n$ we have
$(\phi_*(e_j),\phi_*(X))=(e_j,X)=\phi^*(f^j)(X)
=f^j(\phi_*(X))$ and therefore $\phi_*(e_j)$ is dual to $f^j$.

If $g$ is any function on $N$ we have $n(g\circ\phi)=\phi_*(n)(g)=0$
and $e_j(g\circ\phi)=\phi_*(e_j)(g)=(f_j(g))\circ\phi$.
Thus
\begin{eqnarray*}
   \phi_*([n,e_j])(g)
   &=&[n,e_j](g\circ\phi)=n(e_j(g\circ\phi))-e_j(n(g\circ\phi))\\
   &=&n(f_j(g)\circ\phi)-e_j(n(g\circ\phi))
   =0.
\end{eqnarray*}
Thus $\phi_*([n,e_j])=0$ and hence $([n,e_j],e_k)=0$ for all $j,k=1,2$..

Likewise we see that 
\begin{eqnarray*}
  \phi_*([e_i,e_j])(g)&=&[e_i,e_j](g\circ\phi)
  =e_i(e_j(g\circ\phi))-e_j(e_i(g\circ\phi))\\
  &=&(f_i(f_j(g)))\circ\phi-(f_j(f_i(g)))\circ\phi=[f_i,f_j](g)\circ\phi.
\end{eqnarray*}
Hence $\phi_*([e_i,e_j])=[f_i,f_j]$ and thus $([e_i,e_j],e_k)
=([f_i,f_j],f_k)$ if $i,j,k \in \{ 1, 2\}$.
Since
$$
        (\nabla^M_{e_i}e_j, e_k ) = \sfrac{1}{2} \Bigl(
        ([e_i, e_j], e_k) - (e_i, [e_j, e_k]) + ([e_k, e_i], e_j)\Bigr)
$$
and likewise for the covariant derivatives of the basis $f_1,f_2$
we see that $(\nabla^M_{e_i}e_j, e_k ) = (\nabla^N_{f_i}f_j, f_k ) $.

Since $([n,e_1],e_2)=([n,e_2],e_1)=0$ we have that 
\begin{eqnarray*}
  d\nu(e_1,e_2)&=&\nabla^M_{e_1}\nu(e_2)-\nabla^M_{e_2}\nu(e_1)
  =(\nabla^M_{e_1}n,e_2)-(\nabla^M_{e_2}n,e_1)\\
  &=&(\nabla^M_ne_1,e_2)-(\nabla^M_ne_2,e_1)=2(\nabla^M_ne_1,e_2).
\end{eqnarray*}
and 
$$
   (n,[e_1,e_2])=(n,\nabla^M_{e_1}e_2-\nabla^M_{e_2}e_1)
   =-2(\nabla^M_{e_1}n,e_2)=-2(\nabla^M_ne_1,e_2).
$$
\end{proof}

It is straightforward to check that a
spinor bundle lifts to a spinor bundle under a 
Riemannian submersion as stated in the next proposition.
\begin{prop}[Lifting spinor bundles]\label{prop:liftingspinors}
Let $\phi:M\to N$ be a Riemannian submersion.
If $\Psi_N$ is a spinor bundle on $N$ with 
Clifford map $\sigma_N$ then 
{\it the induced bundle}
$$
    \Psi_M=\phi^*(\Psi_N)=\set{(p,v)\in M\times\Psi_N}{\pi(v)=\phi(p)}
$$
($\pi:\Psi_N\to N$ is the projection map of $\Psi_N$)
is a spinor bundle
on $M$ with 
corresponding Clifford map defined
by
\begin{equation}\label{eq:sigmaM}
    \sigma_M(\phi^*\omega)=\sigma_N(\omega)\circ\phi
\end{equation}
for all one forms $\omega$ on $N$ 
and 
\begin{equation}\label{eq:sigmanu}
    \sigma_M(\nu)=-\left(i\sigma_N(f^1)\sigma_N(f^2)\right)\circ\phi
\end{equation}
if $f^1$, $f^2$ is an oriented orthonormal
frame in $T^*N$.
\end{prop}
%Note that the Hodge dual 
%of $\nu$ is the pull back of the volume form on $N$, 
%i.e, $*\nu=\phi^*(vol_N)$. Using the extension of
%the Clifford multiplication to higher order forms we 
%have that $\sigma_N(f^1)\sigma_N(f^2)=\sigma_N(vol_{N})$.
%In particular we see that (\ref{eq:sigmanu}) is equivalent to
%$\sigma_M(\phi^*(vol_{N}))=\sigma_N(vol_{N})$,
%which is a consequence of (\ref{eq:sigmaM}).

We  recall that there is a natural connection
on an induced bundle.
\begin{prop}[Induced connection]\label{prop:indco}
If $\nabla$ is any connection on 
$\Psi_N$ then there is a unique connection
$\phi^*(\nabla)$ on the induced bundle $\phi^*(\Psi_N)$
such that the chain rule 
$
 \phi^*(\nabla)_Y\left(\xi\circ\phi\right)=\left(\nabla_{\phi_*(Y)}\xi\right)
        \circ \phi
$
is satisfied
for all $Y\in T_*M$ and all sections $\xi$ of $\Psi_N$.
\end{prop}
\begin{proof} This result is standard in the theory of vector bundles.
Locally on $\Psi_N$ we may choose a
basis $\xi_{\pm}$. Then $\xi_{\pm}\circ\phi$ is a local 
basis on $\phi^*(\Psi_N)$. Any section in 
$\phi^*(\Psi_N)$ may locally be written in terms of 
$\xi_{\pm}\circ\phi$. We may use this to extend $\phi^*(\nabla)$
to all sections of $\phi^*(\Psi_N)$. Note that the 
extension is unique. It is straightforward to 
see that this extension locally defines a connection and
that the extension is independent of the choice of the local
basis $\xi_{\pm}$. Hence the connection is globally
defined on $\phi^*(\Psi_N)$ and it satisfies the chain rule.
\end{proof}

If $\nabla^N$ is a $Spin^c$ connection on $\Psi_N$ it is however not 
necessarily true that the induced connection $\phi^*(\nabla^N)$
is a $Spin^c$ connection on $\Psi_M=\phi^*(\Psi_N)$.
It is however easy to correct it such that it becomes a $Spin^c$ 
connection.

\begin{prop}[Lifting $Spin^c$ connections]\label{prop:liftingconnections}
Let again $\phi:M\to N$ be a Riemannian submersion
and $\nu = *\phi^*(vol_N)$. 
Let $\nabla^N$ be a $Spin^c$ connection on the spinor bundle
$\Psi_N$. Let $\phi^*(\nabla^N)$ be the induced connection 
on $\Psi_M=\phi^*(\Psi_N)$.
Then 
\begin{equation}
        \nabla_X^M:=\phi^*(\nabla^N)_X
        -\sfrac{1}{2}\sigma_M(\nu)\sigma_M(\nabla^M_X\nu)
     -\sfrac{i}{4}\nu(X)(\nu,*d\nu)\sigma_M(\nu)
\label{def:nablam}
\end{equation}
is a $Spin^c$ connection on $\Psi_M$.
\end{prop}
\begin{proof} As in the proof of Prop.~\ref{prop:indco}
we see that 
it is enough to check the conditions (\ref{item:spinco1})
and (\ref{item:spinco2}), from the Definition~\ref{defi:spinco}
of $Spin^c$ connections, for spinor fields of the form 
$\xi\circ\phi$.
The first condition (\ref{item:spinco1}) is clear
since $\nabla^N$ is a $Spin^c$ connection.

To check (\ref{item:spinco2}) it is enough to consider 
 either for  the one form $\alpha=\nu$ or any locally defined one form 
of the form $\alpha=\phi^*(\omega)$, where $\omega$ is a locally defined 
one-form  on $N$. 
We begin with the latter case. We
first calculate $\nabla^M_X\phi^*(\omega)$.
As in Prop.~\ref{prop:pullvolform} let $e^1=\phi^*(f^1)$,
$e^2=\phi^*(f^2)$ be local 
one forms such that $\nu,e^1,e^2$ is an orthonormal basis and 
let $n,e_1,e_2$ be the dual basis of vectors.
{F}rom Prop.~\ref{prop:pullvolform}
we know that $(\nabla^M_{e_i}e^j,e^k)=(\nabla^N_{f_i}f^j,f^k)$
for $i,j,k \in \{ 1, 2\}$
and therefore $\phi^*\left[\nabla^N_{f_i}\omega\right]$ 
is the projection of $\nabla^M_{e_i}\phi^*(\omega)$ orthogonal 
to $\nu$.
We can therefore write in general
\begin{eqnarray*}
        \nabla^M_X\phi^*(\omega)
        &=&\nabla^M_{(X,n)n}\phi^*(\omega)+
        \nabla^M_{X-(X,n)n}\phi^*(\omega)\\
        &=&(X,n)\nabla^M_n\phi^*(\omega)+
        \left(\nu,\nabla^M_{X-(X,n)n}\phi^*(\omega)\right)\nu
        +\phi^*\left[\nabla^N_{\phi_*(X)}\omega\right]\\
        &=&\nu(X)\left(\nabla^M_n\phi^*(\omega)-
          \left(\nu,\nabla^M_n\phi^*(\omega)\right)\nu\right)
        -\left(\nabla^M_{X}\nu,\phi^*(\omega)\right)\nu \\ &&{}
        +\phi^*\left[\nabla^N_{\phi_*(X)}\omega\right].
\end{eqnarray*}
Note now that since $(\phi^*(\omega),e^j)=(\omega,f^j)\circ\phi$ is
constant in the direction $n$ we have
$$
    \nabla^M_n\phi^*(\omega)=
    (\phi^*(\omega),e^1)\nabla^M_ne^1+(\phi^*(\omega),e^2)\nabla^M_ne^2.
$$  
Moreover since $\nabla_n e^j$ is orthogonal to $e^j$ we have
\begin{eqnarray*}
    \nabla^M_n\phi^*(\omega)-\left(\nu,\nabla^M_n\phi^*(\omega)\right)\nu&=&
    (\phi^*(\omega),e^1)(\nabla^M_ne^1,e^2)e^2\\{}&&
    +(\phi^*(\omega),e^2)(\nabla^M_ne^2,e^1)e^1\\
    &=&\sfrac{1}{2}(\nu,*d\nu)\left(
        (\phi^*(\omega),e^1)e^2-(\phi^*(\omega),e^2)e^1\right),
    \end{eqnarray*}
using Prop.~\ref{prop:pullvolform}. Thus 
since $\left[\sigma_M(\nu),\sigma_M(e^1)\right]=2i\sigma_M(e^2)$
and $\left[\sigma_M(\nu),\sigma_M(e^2)\right]=-2i\sigma_M(e^1)$
we find that 
\begin{eqnarray}
  \sigma_M\left(\nabla^M_X\phi^*(\omega)\right)&=&
  -\sfrac{i}{4}\nu(X)(\nu,*d\nu)\left[\sigma_M(\nu),\sigma_M
\left(\phi^*(\omega)\right)
  \right]
        -\left(\nabla^M_{X}\nu,\phi^*(\omega)\right)\sigma_M(\nu)
        \nonumber \\ &&{}
        +\sigma_M\left(\phi^*\left[\nabla^N_{\phi_*(X)}\omega\right]
          \right).\label{eq:phi*1}
\end{eqnarray}

On the other hand using (\ref{def:nablam})
\begin{eqnarray*}
 \nabla^M_X[\sigma_M(\phi^*(\omega))\xi\circ\phi]&=&
       \nabla^N_{\phi^*(X)}[\sigma_N(\omega)\xi]\circ\phi\\ &&{}
 -\sfrac{1}{2}\sigma_M(\nu)\sigma_M(\nabla^M_X\nu)
        \sigma_M(\phi^*(\omega))\xi\circ\phi\\ &&{}
     -\sfrac{i}{4}\nu(X)(\nu,*d\nu)\sigma_M(\nu)
        \sigma_M(\phi^*(\omega))\xi\circ\phi.
\end{eqnarray*}
Hence
\begin{eqnarray*}
  \left[\nabla^M_X,\sigma_M(\phi^*(\omega))\right]\xi\circ\phi
  &=& \left(\sigma_N\left(\nabla^N_{\phi^*(X)}\omega\right)\xi\right)
  \circ\phi\\&&{}
  -\sfrac{1}{2}\left[\sigma_M(\nu)\sigma_M(\nabla^M_X\nu),
        \sigma_M(\phi^*(\omega))\right]\xi\circ\phi\\ &&{}
     -\sfrac{i}{4}\nu(X)(\nu,*d\nu)\left[\sigma_M(\nu),
        \sigma_M(\phi^*(\omega))\right]\xi\circ\phi
\end{eqnarray*}
and we have that 
\begin{eqnarray*}
  \left[\nabla^M_X,\sigma_M(\phi^*(\omega))\right]&=&
  \sigma_M\left(\phi^*\left(\nabla^N_{\phi^*(X)}\omega\right)\right)
  -\left(\nabla^M_X\nu,\phi^*(\omega)\right)\sigma_M(\nu)\\
  &&-\sfrac{i}{4}\nu(X)(\nu,*d\nu)\left[\sigma_M(\nu),
        \sigma_M(\phi^*(\omega))\right]=
      \sigma_M\left(\nabla^M_X\phi^*(\omega)\right).
\end{eqnarray*}
The last identity follows from (\ref{eq:phi*1}).

It remains to check that 
$\left[\nabla^M_X,\sigma_M(\nu)\right]=\sigma_M\left(\nabla^M_X\nu\right)$.
This follows from repeated use of the identity just proved
since from (\ref{eq:sigmanu}) we have $\sigma_M(\nu)=-i\sigma_M(\phi^*(f^1))
\sigma_M(\phi^*(f^2))$. Thus if we again write $e^j=\phi^*(f^j)$, 
$j=1,2$ we get 
\begin{eqnarray*}
 \left[\nabla^M_X,\sigma_M(\nu)\right]&=&-i\sigma_M(\nabla^M_Xe^1)
 \sigma_M(e^2)-i\sigma_M(e^1)\sigma_M(\nabla^M_Xe^2)\\
&=&-i(\nabla_X^Me^1,\nu)\sigma_M(\nu)\sigma_M(e^2)
-i(\nabla_X^Me^2,\nu)\sigma_M(e^1)\sigma_M(\nu)\\&&
-i(\nabla_X^Me^1,e^2)-i(\nabla_X^Me^2,e^1)\\
&=&i(e^1,\nabla_X^M\nu)\sigma_M(\nu)\sigma_M(e^2)+
i(e^2,\nabla_X^M\nu)\sigma_M(e^1)\sigma_M(\nu)\\
&=&
\sigma_M(\nabla_X^M\nu),
\end{eqnarray*}
where we have used that $(\nabla_X^Me^1,e^2)+(\nabla_X^Me^2,e^1)=0$
and $i\sigma_M(e^1)\sigma_M(\nu)=\sigma_M(e^2)$ and
$i\sigma_M(\nu)\sigma_M(e^2)=\sigma_M(e^1)$.
\end{proof}

\begin{prop}
  The magnetic 2-form of $\nabla^M$, defined 
by (\ref{def:nablam}) is the pull back of the 
magnetic 2-form of $\nabla^N$.
\end{prop}
\begin{proof}
  Let $\xi_\pm$ be a local basis for $\Psi_N$. Then 
$\xi_\pm\circ\phi$ is  a local basis for $\Psi_M$.
We know from Prop. \ref{prop:local}
 and \ref{prop:magnfield}  that if we define local one-forms
$\alpha_N$ and $\alpha_M$ on $N$ and $M$ 
respectively by
\begin{eqnarray*}
   \alpha_M(X)&=&\sfrac{i}{2}\Big[	
  \left(\xi_+\circ\phi,\nabla_X^M(\xi_+\circ\phi)\right)
  +\left(\xi_-\circ\phi,\nabla_X^M(\xi_-\circ\phi)\right)\Big],
  \quad X\in T_*M
\\
\noalign{\hbox{and}}
 \alpha_N(Y)&=&\sfrac{i}{2}\Big[	(\xi_+,\nabla^N_{Y}\xi_+)
  +(\xi_-,\nabla^N_{Y}\xi_-)\Big],\quad Y\in T_*N
\end{eqnarray*}
then $d\alpha_M$ and $d\alpha_N$ are the magnetic 2-forms
on $M$ and $N$ respectively.
We now show that $\phi^*(\alpha_N)=\alpha_M$ which implies
the statement of the proposition.

Since $\nu$ is orthogonal to $\nabla_X^M\nu$ we have that 
the anticommutator 
$$\{\sigma_M(\nu),\sigma_M(\nabla_X^M\nu)\}=0$$ 
and hence $\sigma_M(\nu)\sigma_M(\nabla_X^M\nu)$ is traceless.
Therefore 
\begin{eqnarray*}
  \alpha_M(X)
  &=&\sfrac{i}{2}\Big[
  \left(\xi_+\circ\phi,\phi^*(\nabla^N)_X(\xi_+\circ\phi)\right)
  +\left(\xi_-\circ\phi,\phi^*(\nabla^N)_X(\xi_-\circ\phi)\right)\Big]\\
  &=&\sfrac{i}{2}\Big[(\xi_+,\nabla^N_{\phi_*(X)}\xi_+)
  +(\xi_-,\nabla^N_{\phi_*(X)}\xi_-)\Big]=\alpha_N(\phi_*(X)).
\end{eqnarray*}
\end{proof}

We now turn to how the Dirac operator lifts. 

\begin{lm}[Lifting the Dirac operator by a Riemannian submersion]
\label{lm:Diraclift}
Let $\phi:M\to N$ be a Riemannian submersion  and let $\nu = *\phi^*(vol_N)$.
 Let $\Psi_N$ be a 
spinor bundle on $N$ with $Spin^c$ connection $\nabla^N$ and 
let $\Psi_M$ and $\nabla^M$ be the lifts described in 
Propositions~\ref{prop:liftingspinors} and
\ref{prop:liftingconnections}. For the corresponding 
Dirac operators ${\cal D}_N$ on $\Psi_N$ and
${\cal D}_M$ on $\Psi_M$ we have for all sections $\xi$ 
in $\Psi_N$ that
\begin{equation}\label{eq:Diracpullback1} 
%   {\cal D}_M(\xi\circ\phi)=\left({\cal D}_N\xi\right)\circ\phi
%   +\sfrac{i}{2}\sigma_M(\nabla^M_{n}\nu)\xi\circ\phi
%     +\sfrac{1}{4}(\nu,*d\nu)\xi\circ\phi.
     {\cal D}_M(\xi\circ\phi)=\left({\cal D}_N\xi\right)\circ\phi
     +\sfrac{1}{2}\sigma_M\left( *d\nu -\sfrac{1}{2}(\nu, *d\nu)\nu\right)
     \sigma_M(\nu)\xi\circ\phi.
\end{equation}
\end{lm}

\begin{proof}
We choose to write the Dirac operator locally
as ${\cal  D}_M=-i\sigma_M(e^1)\nabla^M_{e_1}-i\sigma_M(e^2)\nabla^M_{e_2}
-i\sigma_M(\nu)\nabla^M_{n}$. We find  from Prop.~\ref{prop:liftingconnections}
that 
$$
{\cal D}_M\left(\xi\circ\phi\right)=\left({\cal
     D}_N\xi\right)\circ\phi
+A\xi\circ\phi,
$$
where the matrix $A$ is 
\begin{eqnarray*}
    A&=&\sfrac{i}{2}\sigma_M(e^1)\sigma_M(\nu)\sigma_M(\nabla^M_{e_1}\nu)
    +\sfrac{i}{2}\sigma_M(e^2)\sigma_M(\nu)\sigma_M(\nabla^M_{e_2}\nu)
    \\
    &&+\sfrac{i}{2}\sigma_M(\nabla^M_{n}\nu)-\sfrac{1}{4}(\nu,*d\nu).
\end{eqnarray*}

{F}rom Prop.~\ref{prop:pullvolform}
we have that $(\nabla^M_{e_i}\nu,e^i)=(\nabla^M_{n}e^i,e^i)=0$ for $i=1,2$. 
Since also $(\nabla^M_{e_i}\nu,\nu)=0$ we conclude therfore that 
\begin{eqnarray*}
    \nabla^M_{e_1}\nu&=&(\nabla^M_{e_1}n,e_2)e^2=\sfrac{1}{2}
    (\nu,*d\nu)e^2,
    \\ \noalign{\hbox{and}}
    \nabla^M_{e_2}\nu&=&-(\nabla^M_{e_1}n,e_2)e^1=-\sfrac{1}{2}
    (\nu,*d\nu)e^1.
\end{eqnarray*}
Thus since $\sigma_M(e^1)\sigma_{M}(e^2)\sigma_{M}(\nu)=i$ we have
\begin{eqnarray*}
  A&=&\sfrac{i}{4}(\nu,*d\nu)\left(
    \sigma_{M}(e^1)\sigma_{M}(\nu)\sigma_{M}(e^2)-
    \sigma_{M}(e^2)\sigma_{M}(\nu)\sigma_{M}(e^1)\right) \\
  &&+\sfrac{i}{2}\sigma_M(\nabla^M_{n}\nu)-\sfrac{1}{4}(\nu,*d\nu)
  =\sfrac{i}{2}\sigma_M(\nabla^M_{n}\nu)+\sfrac{1}{4}(\nu,*d\nu) .
\end{eqnarray*}
Hence we just proved that
\begin{equation}
 {\cal D}_M(\xi\circ\phi)=\left({\cal D}_N\xi\right)\circ\phi
  +\sfrac{i}{2}\sigma_M(\nabla^M_{n}\nu)\xi\circ\phi
   +\sfrac{1}{4}(\nu,*d\nu)\xi\circ\phi.
\label{eq:Diracpullback}
\end{equation}

Using  $\sigma_M(e^1)\sigma_M(e^2)\sigma(\nu)=i$
note that
$$
   i\sigma_M(e^1)\xi\circ\phi= \sigma_M(e^2)\sigma_M(\nu)\xi\circ\phi
   \quad \mbox{and}\quad 
   i\sigma_M(e^2)\xi\circ\phi=  -\sigma_M(e^1)\sigma_M(\nu)\xi\circ\phi .
$$
Thus we have that
${\cal D}_M(\xi\circ\phi)-\left({\cal D}_N\xi\right)\circ\phi$ is equal to
$$
   \Big(\sfrac{1}{2}(\nabla^M_{n}\nu,e^1)\sigma_M(e^2)
   -\sfrac{1}{2}(\nabla^M_{n}\nu,e^2)\sigma_M(e^1)+
   \sfrac{1}{4}(\nu,*d\nu)\sigma_M(\nu)\Big)\sigma_M(\nu)\xi\circ\phi.
$$
If we now use that 
$$
  (e^1,*d\nu)=d\nu(e_2,n)=(\nabla_{e_2}^M\nu,\nu)-(\nabla_n^M\nu,e^2)=
  -(\nabla_n^M\nu,e^2)
$$
and likewise that $(e^2,*d\nu)=(\nabla_n^M\nu,e^1)$ we obtain
\ref{eq:Diracpullback1} .
\end{proof}

Of course it would have been nicer if the last term in
(\ref{eq:Diracpullback1}) had not been present. 
One may try to change the connection on $M$, by adding a
one-form, in order to cancel the last term in (\ref{eq:Diracpullback1}).
It is however not possible to cancel this term for all spinors,
but only for spinors with spin pointing in a specific direction.
The following theorem is a simple consequence of Lemma~\ref{lm:Diraclift}.
\begin{thm}\label{thm:pullup} We use the notation of Lemma~\ref{lm:Diraclift}.
If we therefore define new $Spin^c$ connections 
\begin{equation}\label{eq:nablapm}
  \nabla^{M,\pm}:=\nabla^M \mp \sfrac{i}{2} \left(*d\nu
 -\sfrac{1}{2}(\nu,*d\nu)\nu\right)
\end{equation}
we obtain for the corresponding Dirac operators ${\cal D}^\pm_M$ 
that 
\begin{equation}\label{eq:diracpm}
  {\cal D}^\pm_M(\xi\circ\phi)=\left({\cal D}_N\xi\right)\circ\phi,
\end{equation}
for spinors satisfying $\sigma_M(\nu)\xi\circ\phi=\pm\xi\circ\phi$.
In particular, if $\xi \in \Psi_N$ is a normalized
zero mode of ${\cal D}_N$ with
a definite spin direction 
$$
	S = \left(
\xi, -i\sigma_N(f^1)\sigma_N(f^2)\xi\right)
\in \{ + , - \},
$$
 then $\xi\circ\phi \in \Psi_M$
is a zero mode of ${\cal D}_M^{sgn (S)}$.
\end{thm}

\section{Lower bound on the number of zero modes}\label{sec:lowerbound}

Theorem~\ref{thm:pullup} allows us to construct zero modes of Dirac
operators on $M$ with a certain connection starting from definite-spin
zero modes 
of a Dirac operator ${\cal D}_N$ on $N$. 
On the other hand, 
the index theorem for  ${\mathcal D}_N$ claims that 
$$
        \hbox{ind} \; {\mathcal D}_N=\frac{1}{2\pi}\int_N\beta_N,
$$
where $\beta_N$ is the magnetic two-form of $\nabla_N$.
This integer is the Chern number of  
the determinant line bundle of $\Psi_N$. 
The index is defined as
\begin{eqnarray*}
        \hbox{ind} \; {\mathcal D}_N
        &=&\dim\set{\xi\in\Gamma(\Psi_N)}{{\mathcal D}_N\xi=0,\ \sigma_M(\nu)\xi=   
        \xi}\\ &&{}
-\dim\set{\xi\in\Gamma(\Psi_N)}{{\mathcal D}_N\xi=0,\ \sigma_M(\nu)\xi=-
\xi} \; ,
\end{eqnarray*}
where we have abused notations and let $\sigma_M(\nu)$ 
act on sections in $\Psi_N$. The action is of course the one given in 
(\ref{eq:sigmanu}).

Theorem~\ref{thm:pullup} and the index Theorem
together give
the following theorem.

\begin{thm}[Lower bound on the number of zero modes]\label{thm:lowerbound}
Let $\Psi_N$ be a spinor bundle over a 2-dimensional manifold $N$,
let $\nabla^N$ be a $Spin^c$ connection and let ${\cal D}_N$ be
the Dirac operator.  We denote 
the magnetic two-form of $\nabla_N$ by $\beta_N$ and let 
$$
\Phi: = 
\frac{1}{2\pi}\int_N\beta_N \quad \mbox{and}\quad  s : =\mbox{sgn}(\Phi) 
	\in \{ + , -\} .
$$
Consider a Riemannian submersion $\phi: M \to N$ from some 3-dimensional
manifold $M$ and let $\Psi_M$ and $\nabla^M$ be the lifts
of the spinor bundle and the connection as described in
Propositions~\ref{prop:liftingspinors} and
\ref{prop:liftingconnections}. Let $\nu := *\phi^*(vol_N)$
and define a new  $Spin^c$ connection
$$
	\widetilde\nabla^{M}: = \nabla^M - i\sfrac{s}{2}
	\left(*d\nu -\sfrac{1}{2}(\nu,*d\nu)\nu\right)
$$
on $\Psi_M$. The corresponding Dirac
operator is denoted by $\widetilde{\cal D}_M = (-i)\sigma(\wt \nabla^M)$.
Then
\begin{equation}
	\mbox{dim}\; \mbox{Ker} \; \widetilde{\cal D}_M
	\ge \Big| {1\over 2\pi}\int_N \beta_N\Big|
\label{eq:lowerbound}
\end{equation}
\end{thm}

%\begin{remark} Depending on the sign of $\int_N \beta_N$, this
%theorem gives a nontrivial lower bound on the number
%of linearly independent zero modes {\it either} for  ${\cal D}_M^+$
%{\it or} for ${\cal D}_M^-$. We do not have information about the
%other operator in terms of $\beta_N$. 
%\end{remark}

\begin{remark} The magnetic two-form of $\widetilde{\cal D}_M$ is 
$$
	\beta_M : = \phi^*(\beta_N) + \sfrac{s}{2}
	\; d\left( *d\nu - \sfrac{1}{2} (\nu, *d\nu)\nu\right) .
$$
We do not have an independent characterization of all magnetic
two-forms $\beta$ on $M$ which can be presented in this form.
\end{remark}

\begin{remark} It is very interesting to investigate the cases
of equality in (\ref{eq:lowerbound}). Our method loses equality
in two different steps. Less seriously, we estimate the dimension of 
the kernel of ${\cal D}_N$ by the index, which is not sharp
unless some vanishing theorem holds for ${\cal D}_N$.
 More importantly, we only considered 
those zero modes of $\widetilde{\cal D}_M$ whose spin direction
is parallel with $\nu$. We will show in the next section
that in the special case of the Hopf map $\phi: \S^3\to \S^2$
we have equality in (\ref{eq:lowerbound}).
\end{remark}

{\it Proof of Theorem \ref{thm:lowerbound}}.
We  recall that
$\widetilde\sigma:
=-i\sigma_N(f^1)\sigma_N(f^2)$ commutes with ${\cal D}_N$. Let
${\cal D}_N^\pm$ be the restriction of ${\cal D}_N$ onto
the subspaces $\widetilde\sigma\xi =\pm\xi$. Then by the index theorem
$$
	\mbox{dim}\; \mbox{Ker} \; {\cal D}_N^+
 -\mbox{dim}\; \mbox{Ker} \; {\cal D}_N^- = \Phi= {1\over 2\pi}\int_N \beta_N ,
$$
which implies that
$$
	\max_\pm \mbox{dim}\; \mbox{Ker} \; {\cal D}_N^\pm \ge |\Phi| .
$$
Notice that $ \widetilde{\cal D}_M = {\cal D}_M^s$, where the
sign superscript is $s = \mbox{sgn} (\Phi) \in \{ +, - \}$.
Hence,
using Theorem \ref{thm:pullup} we have
$$
	\mbox{dim}\; \mbox{Ker} \; \widetilde{\cal D}_M
	\ge \max_\pm \mbox{dim}\; \mbox{Ker} \; {\cal D}_N^\pm ,
$$
which completes the proof.
\qed

\section{Geometry of the Hopf map}\label{sec:hopf}

We identify $\S^2$ with the Riemann sphere $\bar \C= \C\cup \{ \infty\}$.
The standard metric on $\S^2$ can be written as $g_2 =
 \left( 1 + \sfrac{1}{4} |w|^2\right)^{-2}dw\, d\bar w$.
The 3-sphere we write as
 $\S^3 = \{ (z_1, z_2) \; : \; |z_1|^2 + |z_2|^2 =1\}\subset \C^2$
and let $g_3$ be the standard metric on $\S^3$. The Hopf map
$\phi : \S^3\to \S^2$ can then be written as $\phi(z_1, z_2)= 2z_1z_2^{-1}$.
We remark, that 
for our purposes the conjugate map 
$\phi'(z_1, z_2)= 2\bar z_1\bar z_2^{-1}$ could have been chosen as well.
We summarize a few geometric  properties of the Hopf map.
\begin{lm}[Hopf map is Riemannian submersion]
The Hopf map is a Riemannian submersion between
the Riemannian manifolds $M= (\S^3, g_3)$ and $N= (\S^2, \sfrac{1}{4}g_2)$.
Let $\nu= * \phi^*(vol_N)$ and let $n$ be the vector field corresponding
to the one-form $\nu$. Then $n$ is a geodesic vectorfield on $\S^3$,
its integral curves are main circles of $\S^3$. Moreover
we have the relation
\begin{equation}
	d\nu = - 2*\!\nu .
\label{eq:dmu}
\ee
\end{lm}

\begin{proof} 
The vectorfield $v := (z_1, z_2)$ on $\C^2$ is orthogonal to $\S^3$.
Define the following three vectorfields on $\C^2$; 
$u_1: = (i\bar z_2, -i\bar z_1)$, $u_2:=(\bar z_2, -\bar z_1)$
and $n : = (iz_1, iz_2)$. It is easy to
see that they are orthogonal to $v$, hence they are
 also  vectorfields on $\S^3$. Moreover, they form a positively
oriented orthonormal basis of $T_*\S^3$.

The integral curves 
$\chi(t) = e^{it} n$ of the vectorfield $n$
are  main circles in $\S^3$. Hence $n$ is a geodesic vectorfield on $\S^3$.
Moreover $\phi\circ \chi(t)$ is independent of $t$, so  $\phi_* (n) =0$
and $\phi$ 
is a fibration with $\S^1$ fibers, all having the same length $2\pi$.

To check that $\phi:M\to N$ is a partial isometry, we compute
the pushforwards of $u_1, u_2$ at any point $(z_1, z_2)$:
$$
	\phi_*(u_1) = \left( i\bar z_2 \frac{\partial}{\partial z_1}
	-i z_2 \frac{\partial}{\partial \bar z_1} 
	-i\bar z_1  \frac{\partial}{\partial z_2}
	+i z_1 \frac{\partial}{\partial \bar z_2}\right)
	\frac{2z_1}{z_2} =\frac{2i}{z_2^2}
$$
and similarly $\phi_*(u_2)= 2z_2^{-2}$. It is clear that $\phi_*(u_1)$
and $\phi_*(u_2)$ are orthogonal and we check that their length is
$$
	\frac{1}{2} \left( 1 + \frac{1}{4} |\phi(z_1, z_2)|^2\right)^{-1} 
	\frac{2}{|z_2|^2}
	= 1.
$$

Moreover, $vol_N = f^1\wedge f^2$, where $f^j$ is the
the one-form on $\S^2$ dual to $\phi_*(u_j)$, hence $\nu = *\phi^*(vol_N)$
is dual to the vector $n$.

In order to compute $*d\nu$, we first compute
 the Levi-Civita connection on $\S^3$ 
in the basis $u_1, u_2, n$. For example
$$
	\nabla_n u_1 = p\left( iz_1 \frac{\partial}{\partial z_1}
	-i\bar z_1 \frac{\partial}{\partial \bar z_1} + 
	iz_2  \frac{\partial}{\partial z_2}
	-i\bar z_2 \frac{\partial}{\partial \bar z_2}\right)(i\bar z_2, \;
	-i\bar
	z_1) = p( u_2) = u_2,
$$
where $p: T_*\C^2\to T_*\S^3$ is the canonical projection.
Similarly, we obtain  
\be
	\nabla_n u_2= -u_1, \qquad \nabla_{u_1} n = - u_2, \qquad
	\nabla_{u_2} n = u_1,
\label{eq:nablas}
\ee
$$ 
	\nabla_{u_1}u_2= n, \qquad \nabla_{u_2}u_1 = -n, \qquad
	\nabla_nn=0.
$$
Finally, we obtain
$$
	(\nu, *d\nu) = d\nu( u_1, u_2) = (\nabla_{u_1}n, u_2)
	-  (\nabla_{u_2}n, u_1) = -2
$$
and
$$
	(u_1, *d\nu) = d\nu(u_2, n) = (\nabla_{u_2}n , n)-
	(\nabla_n n, u_2) =0
$$
$$
	(u_2, *d\nu) = - d\nu(u_1, n) = - (\nabla_{u_1}n , n) +
	(\nabla_n n, u_1) =0,
$$
and (\ref{eq:dmu}) follows.
\end{proof}

\section{Dirac operators on $\S^3$ and $\R^3$}\label{sec:exact}

We show that in certain cases the only zero modes of a
three dimensional Dirac operator are those which were
obtained in Theorem \ref{thm:pullup}. Moreover, for 
these operators we are able to determine the full spectrum.

\subsection{Spinor bundles on $\S^3$ and $\S^2$}

We discuss $Spin^c$ spinor bundles on $\S^2$ and $\S^3$
in the Appendix. Here we just mention that
on $\S^3$ there is only one (up to isomorphisms) 
$Spin^c$ bundle and it is just a trivial bundle.
 For any closed 2-form $\beta$ there is a 1-form $\alpha$
with $d\alpha = \beta$ since $H^2(\S^3, \R)=\{ 0\}$. Hence any  $\beta$ is the 
magnetic 2-form of some $Spin^c$ connection. This connection is given
for example by the formula in Proposition \ref{prop:local}.

Moreover, the connection is up to an overall gauge transformation 
uniquely determined by the magnetic 2-form, in particular
the  spectrum of the Dirac operator depends only on the magnetic 
2-form. Indeed, 
if $\nabla^{M, 1}$ and $\nabla^{M, 2}$ both have magnetic
two form $\beta$, then by Proposition
\ref{prop:unique} $\nabla^{M, 1} -\nabla^{M,2} = i\omega$ with
$d\omega=0$. Since $H^1(\S^3, \R)=0$,  $\omega = df$ with some
function $f\in C^\infty(M)$. Therefore 
$U \nabla^{M,1} U^*= \nabla^{M,2}$ with the unitary operator $U=e^{if}$,
and the same
relation holds for the corresponding Dirac operators. The spectrum
is gauge invariant.

On $\S^2$ there are inequivalent $Spin^c$ bundles.
 For each $n\in\Z$ there is a $Spin^c$ bundle
$\Psi_n$ on $\S^2$, such that all $Spin^c$ connections on $\Psi_n$
have the property that the corresponding magnetic 2-form
integrated over $\S^2$ gives $2\pi n$
 (Sections~\ref{sec:spinons2}, \ref{sec:spincons2}).
The bundles $\Psi_n$ exhaust all $Spin^c$ bundles on $\S^2$
(Proposition~\ref{prop:isom}).
The integer $n$ is the Chern number of $\Psi_n$ (or rather
that of the determinant line bundle of $\Psi_n$).
Moreover any 2-form on $\S^2$ that  integrates to $2\pi n$ for some
$n\in \Z$  is
the magnetic 2-form for some  connection on $\Psi_n$
(Proposition~\ref{prop:anyfield}).
Again the connection is uniquely determined by this  2-form,
up to an overall gauge freedom
and the spectrum of the Dirac operator depends 
only on the magnetic 2-form.

\subsection{Spectrum of the Dirac operator on $\S^3$}

 Let $M= (\S^3, g_3)$, $N= (\S^2, \sfrac{1}{4} g_2)$ and $\phi$ be
the Hopf map, $\nu : = *\phi^*(vol_N)$. Let $\beta_M$ be  a
two-form (magnetic field) on $M$ such that
 $\beta_M= h  *\!\nu$, $h\in C^{\infty}(M)$, i.e. the magnetic field is
parallel with the pull-up volume form. 
Clearly $\beta_M = \phi^*(g \, vol_N)$ for some function $g
\in C^\infty(N)$, $h=g\circ\phi$ (see Remark \ref{rem:2}).

Let $\Psi_M$ be
the spinor bundle on $M$, which is unique up to isomorphism,
and let $\nabla^M$ be a  $Spin^c$ connection on  $\Psi_M$
with magnetic two form $\beta_M$. The corresponding Dirac operator
is  ${\cal D}_M = -i\sigma(\nabla^M)$.
Recall that $\beta_M$
determines ${\cal D}_M $ up to a gauge transformation. 

\begin{thm}[Spectrum of ${\cal D}_M $]\label{thm:exact}
We define
$$
	c : = \Big\langle {1\over (2\pi)^2} \int_M\beta_M\wedge \nu
	\Big\rangle \qquad \mbox{and}\qquad  m: = \Big[  
	{1\over (2\pi)^2} \int_M\beta_M\wedge \nu
	\Big] 
$$
where for any $x\in \R$ we let  $\langle x \rangle$ denote the unique number 
 in $(-\sfrac{1}{2},\sfrac{1}{2}]$ such that $x- \langle x \rangle \in \Z$.
Let $[x] :  = x- \langle x\rangle$ be this integer.

For any $k\in \Z$ let
\begin{equation}
	\beta_N(k): = \left( g- 2(c+k)\right) (vol_N) .
\label{def:betank}
\end{equation}
Then clearly (see Remark \ref{rem:2})
\begin{equation}
	\frac{1}{2\pi}\int_N \beta_N(k) = m - k \in \Z ,
\label{eq:betanflux}
\end{equation}
hence on the spinor bundle $\Psi_{m-k}$ with Chern number $m-k$ on $N$
there exists a two dimensional Dirac operator ${\cal D}_{N, (k)}$ with
magnetic two-form $\beta_N(k)$.
Let $\Sigma_+(k)$ be the positive spectrum of  ${\cal D}_{N, (k)}$ .

(i.) The spectrum of ${\cal D}_M$ is given
\be
	\mbox{Spec} \; {\cal D}_M
	=   \bigcup_{k\in\Z} \Bigg( S_k \cup 
	\Big\{ \pm\sqrt{\lambda^2 +(k+c)^2} -\frac{1}{2}
	\; : \; \lambda\in \Sigma_+(k)\Big\} \Bigg)
\label{eq:dmspec}
\ee
where
$$
        S_k=\left\{
\begin{array}{ll}
        \{k+c-\frac{1}{2}\},&\hbox{if } m>k\\
        \emptyset,&\hbox{if }m=k\\
 \{-k-c-\frac{1}{2}\},&\hbox{if }m<k \; .
\end{array}\right. 
$$

(ii.) The multiplicity of an eigenvalue of ${\mathcal D}_M$
is equal to the number of ways it can be written as 
$\pm\sqrt{\lambda^2 +(k+c)^2} -\frac{1}{2}$ with 
$k\in\Z$ and $\lambda\in \Sigma_+(k)$ counted with 
multiplicity or as an element in $S_k$ counted with multiplicity
$|m-k|$.

(iii.) The eigenspace of ${\cal D}_M$ with eigenvalue in 
$S_k$ contains spinors with definite spin value $\mbox{sgn}(m-k)$,
(i.e, the eigenvalue of $\sigma(\nu)$).
\end{thm}

\begin{remark}\label{rem:2} From the assumption $\beta_M = h *\!\nu$ 
it follows that $\beta_M = \phi^*(g \; vol_N)$ 
for some function $g\in C^\infty (N)$
and that
\be
	{1\over (2\pi)^2}\int_M \beta_M\wedge\nu 
	= {1\over 2\pi}\int_N g \; vol_N .
\label{eq:fubini}
\ee
To see this, we compute $0 = d\beta_M = dh\wedge *\nu = n(h)\; vol_M$
using that $d*\!\nu = d\Big( \phi^*(vol_N)\Big)= 0$ 
from (\ref{eq:dmu}). Hence $h$ is
constant along the Hopf fibers, $h=g\circ\phi$, 
 and therefore $\beta_M$ is a pullback
of a (not necessarily exact) two form on $N$.
 The integral relation (\ref{eq:fubini}) follows 
from using Fubini's theorem in local patches
 and the fact
that the length of the fibers is $2\pi$.
{F}rom this relation   (\ref{eq:betanflux}) is straight forward.
(Recall that $\int vol_N = \pi$).
\end{remark}

The eigenvalues in $S_k$ correspond to the
zero eigenvalues of the 2-dimensional operator ${\mathcal D}_{N, (k)}$.
For these we have used the following 
theorem of Aharonov and Casher \cite{AC}
to say exactly what the 
multiplicity is and exactly what the spin is of the eigenfunctions.
For completeness we give a proof of this theorem in the Appendix. 

\begin{thm}[Aharonov Casher theorem on $\S^2$]\label{thm:AC}
Let $\nabla$ be a covariant 
derivative on the spinor bundle $\Psi_n$ on $\S^2$ with Chern number
$n={1\over 2\pi}\int \beta$, where $\beta$ is the corresponding 2-form.
Let ${\mathcal D}$ be the corresponding Dirac operator. 
Then the dimension of the 
space of harmonic spinors ker$\,\, {\mathcal D}$ is $|n|$.
Moreover,  ker$\,\, {\mathcal D}\subset\{\eta\ |\ 
\sigma(\nu)\eta=\eta\}$ if $n>0$ and  ker$\,\, {\mathcal D}\subset\{\eta\ |\ 
\sigma(\nu)\eta=-\eta\}$ if $n<0$, 
where as before we have written 
$\sigma(\nu)=-i\sigma(f^1)\sigma(f^2)$ for any 
positively oriented (local) orthonormal basis $f^1,f^2$ of one-forms.
$\sigma(\nu)$ is independent of the choice of one-form basis.
\end{thm}

\begin{remark} With the notation of Theorem~\ref{thm:exact}
we see immediately that
for the very special case of the eigenvalue
in $S_0$ we can say that the multiplicity is precisely
$|m|$. If $|c|<1/2$ we cannot say this for the eigenvalues in 
$S_k$ for $k\ne0$ because it is possible that these could
also be written as $\pm\sqrt{\lambda^2 +(k'+c)^2} -\frac{1}{2}$
for some $k'\ne k$ and $\lambda\ne0$.
The case when $c=\sfrac{1}{2}$ 
 is of most interest to us and we formulate it as a theorem.
\end{remark}

\begin{thm}[Dimension of the space of harmonic spinors
 on $\S^3$]\label{thm:zeromodes}
With the notation of Theorem~\ref{thm:exact}
we consider the case when $c=1/2$. In this case the eigenvalue
0 has multiplicity $m$ if $m>0$ and $-m-1$ if $m<-1$ otherwise 
$0$ is not an eigenvalue.
The eigenvalue $-1$ has multiplicity $-m$ if $m<0$ 
and  $m+1$ if $m>-1$.
\end{thm} 
\begin{proof}
This is a simple consequence of Theorem~\ref{thm:exact}.
The eigenvalue in $S_0$ has multiplicity precisely 
$|m|$ and the eigenvalue in $S_{-1}$ has eigenvalue precisely
$|m+1|$. The theorem now follows from considering the different
cases.
\end{proof}

\begin{remark}
Thus for any natural number we can construct a Dirac operator on $\S^3$ 
such that the dimension of the kernel is that given number.
Using the 
conformal invariance of the dimension
of the kernel one may arrive at a construction on $\R^3$
which is summarized below.
One has to discuss the behavior at infinity, but this
is not very difficult.
\end{remark}

\begin{thm}[Zero modes on $\R^3$]\label{thm:R3}
Let $\tau : \R^3\to  M\setminus \{ p \} $ be
the inverse of the stereographic projection onto $\R^3$
from the sphere $M = (\S^3, g_3)$ with a point removed.
Let $\beta=\tau^{*}(h *\!\nu) = (h\circ \tau)\tau^{*}(*\nu)$
 be the pullback of an arbitrary closed two form $h  *\!\nu$ on $\S^3$.
[In other words, $\beta = \wt h \,\, \tau^{*}(*\nu)$ is an arbitrary
closed two form on $\R^3$ which is parallel with $\tau^{*}(*\nu)$
and the function $\wt h\circ\tau^{-1}$ extends to a regular function $h$ on
$\S^3$.]
Then $h=g\circ\phi$ with some function $g\in C^\infty(\S^2)$
and $(\tau^{-1})^* \beta$ extends to a two form $\beta_M=
 (g\circ \phi) *\!\nu$ on $M$.
Let ${\cal D}_{\R^3}$ be the Dirac operator on $\R^3$ with
magnetic two form $\beta$ and let  ${\cal D}_M$ 
be the Dirac operator on $M$ with magnetic
two form $\beta_M$.
 Then
$$
	\mbox{dim Ker}  \; {\cal D} = \mbox{dim Ker} \; {\cal D}_M \; ,
$$
and  $\mbox{dim Ker} \; {\cal D}_M$ is given in Theorem \ref{thm:zeromodes}.
\end{thm}

\begin{proof}  
{F}rom Remark \ref{rem:2}, we easily
see that any two form  
of the form $\beta = \tau^{*}(h \, *\nu)$ is closed if and only if 
$ h=g\circ\phi$ with some function $g$ on $\S^2$.
Then $\beta_M$ defined as $(g\circ \phi) *\!\nu$ on $M$ coincides
with $(\tau^{-1})^* \beta$ on $M\setminus \{ p \}$.

Next,  we recall that the magnetic two form
determines the Dirac operator on $\S^3$ and $\R^3$
 up to bundle isomorphism and
gauge transformation, hence ${\cal D}_{\R^3}$ and  ${\cal D}_M$
in the theorem are well defined. They can be considered
acting on the trivial bundle $\Psi_{\R^3} = \R^3\times \C^2$
and $\Psi_M = M\times \C^2$, respectively.

The stereographic projection isometrically
  identifies $\big( M\setminus \{ p \}, g_3\big)$
 with $(\R^3, \Omega^2 ds^2)$, where
$\Omega (x): = (1+x^2)^{-1}$.
For any normalized spinor $\xi \in \mbox{Ker} \; {\cal D}_M$, $ {\cal D}_M
\xi =0$, we have
${\cal D}_{\R^3} (\Omega \xi) = 0$  by Theorem \ref{thm:conformaldirac}.
By elliptic regularity of ${\cal D}_M$, $\xi$ is smooth, in
particular $\langle \xi, \xi \rangle$ is bounded. Hence
$$	
	\int_{\R^3} \langle \Omega \xi,\Omega \xi \rangle (vol_{\R^3})
	\leq \max_M \langle \xi, \xi \rangle \int_{\R^3} \Omega^2
	(vol_{\R^3}) <\infty
$$
hence $\Omega\xi\in \mbox{Ker} \; {\cal D}_{\R^3}$.

Conversely, if $\psi \in \mbox{Ker} \; {\cal D}_{\R^3}$, then
$ {\cal D}_M \xi =0$ away from $p$ with 
$\xi:=\Omega^{-1}\psi$. It follows from  $\psi\in L^2 (\Psi_{\R^3})$
that $\xi \in L^2(\Psi_M)$, since $\Omega\leq 1$.
These two facts easily imply that
$\xi$ extends to $p$ and $ {\cal D}_M \xi =0$ everywhere.
In fact, consider a sequence of bounded
cutoff functions $\chi_n \in C^\infty(M)$,
$\chi_n(p)=0$, $\chi_n\to 1$ such that $\| d\chi_n \|_{L^2(\Lambda^1(M))}
  \to 0$. Such sequence exists in three dimensions.
 Clearly ${\cal D}_M (\chi_n \xi) \to {\cal D}_M \xi$
and ${\cal D}_M (\chi_n \xi) = -i \sigma(d\chi_n) \xi \to 0$
as $n\to 0$, hence
${\cal D}_M \xi =0$ on $M$ in the sense of distributions.
It  then follows from elliptic regularity of ${\cal D}_M$
and $\xi \in L^2(\Psi_M)$ that $\xi$ is smooth on $M$ and
$\xi\in \mbox{Ker} \; {\cal D}_M$
\end{proof}

\begin{remark} Finally one may note that 
the first zero mode constructed by Loss and Yau in \cite{LY}
is the stereographic projection of 
the one one gets according to Theorem~\ref{thm:zeromodes}
with $m=1$ and $c=1/2$. This corresponds to choosing $g=3$ 
in Theorem~\ref{thm:exact}.

\end{remark}

The proof of Theorem \ref{thm:exact} is divided into subsections.

\subsection{Rotationally symmetric eigenbasis}

 By the properties of $\phi$ and our
assumption on $\beta_M$ the rotation along the $\S^1$ fibers is a symmetry
of the data, hence the generator of this rotation should commute
with ${\cal D}_M$.

\begin{prop}\label{prop:commute}
Let $n$ be the vector dual to $\nu$ and let us define
\begin{equation}
	Q : = -i\nabla_n^M - \frac{1}{2}\sigma(\nu),
\label{def:Q}
\end{equation}
which is symmetric since $\mbox{div} \, \, n = 0$.
Then
\begin{equation}
	\left[ {\cal D}_M,  Q\right] = 0
\label{eq:comm}
\end{equation}
(when acting on smooth sections),
and we also have that
\begin{equation}
	\left\{ {\cal D}_M, \sigma(\nu)\right\} = 2 Q-\sigma(\nu) 
\label{eq:anticomm}
\end{equation}
(on the domain of ${\cal D}_M$).
\end{prop}

\begin{proof}
 We use the basis $u_1, u_2, n$
constructed in Section \ref{sec:hopf}, and let $u^1, u^2, \nu$ be the
dual basis. Sometimes we use the notation $u^3=\nu$, $u_3=n$ for brevity.
We also drop the superscript $M$ from $\nabla^{M}$.
Then
\begin{eqnarray*}
	\left\{ {\cal D}_M, \sigma(\nu)\right\} 
	&= & (-i)\sum_{j=1}^3
	\left\{ \sigma(u^j)\nabla_{u_j}, \; \sigma(u^3)\right\} \\
	&=&
	 (-i)\sum_{j=1}^3\left\{ \sigma(u^j), \sigma(u^3)\right\}
	\nabla_{u_j}
	+ (-i)\sum_{j=1}^3 \sigma(u^j)
	\left[ \nabla_{u_j},  \sigma(u^3)\right]
	\\
	&=& -2i\nabla_n
	 -i\sum_{j=1}^3 \sigma(u^j)\sigma( \nabla_{u_j} u^3)
	= -2i\nabla_n  - 2 \sigma(\nu) = 2 Q- \sigma(\nu) .
\end{eqnarray*}
In the last step we used $-i\sigma (u^1)\sigma(u^2) =\sigma(u^3)$
and various $\nabla_{u_i}u^j$'s computed in (\ref{eq:nablas}).

To prove \ref{eq:comm}, we first compute
\begin{eqnarray*}
	\left[ {\cal D}_M, \sigma(\nu)\right]
	&=& (-i)\sum_{j=1}^3
	\left[ \sigma(u^j)\nabla_{u_j}, \sigma(u^3)\right] \\
	&=&  (-i)\sum_{j=1}^3\left[ \sigma(u^j), \sigma(u^3)\right]
	\nabla_{u_j}
	+ (-i)\sum_{j=1}^3 \sigma(u^j)\left[ 
	\nabla_{u_j},  \sigma(u^3)\right]
	\\
	&=&  2\sigma(u^1)\nabla_{u_2}
	-  2\sigma(u^2)\nabla_{u_1} -2 \sigma(\nu) .
\end{eqnarray*}
Next, we compute
\begin{eqnarray*}
	\left[ {\cal D}_M, \nabla_{n}\right]
	&=& (-i)\sum_{j=1}^2 \sigma(u^j)\left[  \nabla_{u_j},
	\nabla_{n}\right]
	+ i\sum_{j=1}^2 \left[\nabla_{n},
	  \sigma(u^j)\right] \nabla_{u_j}
	\\
	&=& (-i)\sum_{j=1}^2 \sigma(u^j)\left[  \nabla_{u_j},
	\nabla_{n}\right] + 
	i\sigma(u^2)\nabla_{u_1}-  i\sigma(u^1)\nabla_{u_2} .
\end{eqnarray*}
To compute the commutators, we use
$$
	\left[  \nabla_{u_j}, \nabla_{n}\right]
	= R_\Psi (u_j, n) + \nabla_{[u_j, n]}
$$
and Theorem \ref{thm:spincurv} to express $R_\Psi$ in terms
of the Riemannian curvature $R$ and the magnetic two form $\beta_M$.
Then $R$ can be computed from $\nabla_{u_i}u_j$'s (\ref{eq:nablas}),
and the result is
\begin{eqnarray*}
	R(u_1, u_3)u_1 = -u_3 & \quad R(u_1, u_3)u_2 = 0 & \quad
	R(u_1, u_3)u_3 = u_1 \\
	R(u_2, u_3)u_1 = 0 & \quad R(u_2, u_3)u_2 = -u_3 & \quad
	R(u_2, u_3)u_3 = u_2 .
\end{eqnarray*}
By the assumption on the magnetic field in
Theorem \ref{thm:exact}, $\beta_M (u_j, u_3) =0$ for $j=1, 2$. We obtain
$$
	R_\Psi (u_1, u_3) = -\sfrac{i}{2} \sigma(u^2) \; ,
 	\qquad R_\Psi (u_2, u_3) = \sfrac{i}{2} \sigma(u^1) ,
$$
hence  using $[u_1, u_3] = -2u_2$, $[u_2, u_3] =2u_1$ we conclude that
$$
	(-i)\sum_{j=1}^2 \sigma(u^j)\left[  \nabla_{u_j},
	\nabla_{n}\right]  =  2i\sigma(u^1)\nabla_{u_2}
	-2i\sigma(u^2)\nabla_{u_1}  -i\sigma(u^3) .
$$
Therefore
$$	
	\left[ {\cal D}_M, \nabla_{n}\right]
	= i\sigma(u^1)\nabla_{u_2}
	-i\sigma(u^2)\nabla_{u_1}  -i\sigma(u^3) .
$$
and combining this with $\left[ {\cal D}_M, \sigma(\nu)\right]$
computed above, we arrive at (\ref{eq:comm}).
\end{proof}

Since $ {\cal D}_M$ has a pure point spectrum, (\ref{eq:comm})
implies that it has an eigenbasis consisting of eigenspinors
of $Q$. One expects that these eigenspinors 
are actually pull-ups of some spinors in an appropriate spinor bundle
on $N$. This is correct after a gauge transformation which we describe now.

For any $k\in \Z$
let us fix a spinor connection $\nabla^N = \nabla^{N, (k)}$
 with a magnetic form
$\beta_N(k)$ on the spinor bundle $\Psi_{m-k}$ with Chern number $m-k$
on $N$. We  identify $\Psi_M$ with the lift of  $\Psi_{m-k}$.
Let $\nabla^{M,(k)}$ be the lift of $\nabla^{N, (k)}$
according to Proposition \ref{prop:liftingconnections}:
\begin{equation}
	\nabla^{M,(k)}_X : =
	 \phi^*(\nabla^{N, (k)})_X -\frac{1}{2}
	\sigma(\nu)\sigma(\nabla_X\nu) + \frac{i}{2}\nu(X)\sigma(\nu) ,
\label{def:nablamn}
\end{equation}
where we also used (\ref{eq:dmu}).
Let $\beta_{M,(k)} : =
 \phi^* (\beta_N(k))$ be the magnetic form of $\nabla^{M,(k)}_X$.
Finally we  define 
\begin{equation}
	 \widetilde\nabla^{M, (k)}:
	 = \nabla^{M,(k)} + i(c+k)\nu.
\label{def:tildenablam}
\end{equation}
The magnetic two form of $\widetilde\nabla^{M, (k)}$ is 
\begin{eqnarray*}
	\beta_{M,(k)} - (c+k) d\nu
	& =& \beta_{M,(k)} +2(c+k) (*\nu) = \phi^*\Big(
	\beta_N(k) +2 (c+k) (vol_N)\Big) \\
	&=&  \phi^*\Big( g (vol_N)\Big)  =\beta_M
\end{eqnarray*}
by (\ref{def:betank}).
It is now clear that  $\widetilde\nabla^{M, (k)}$ 
and $\nabla^M$ are gauge equivalent
since their magnetic fields are the same.
Therefore there exists a function $f_k\in C^\infty(M)$, depending on $k$,
 such that
$$
	\widetilde\nabla^{M, (k)}  = e^{if_k} \nabla^M e^{-if_k} .
$$
Hence the spectrum of
${\cal D}_M$ and $\widetilde{\cal D}_{M, (k)}=-i\sigma(\wt\nabla^{M, (k)})$
 are the same for any $k\in \Z$,
 and we will work with the latter operators.

Let
$$
	\widetilde Q^{(k)} : = e^{if_k}Qe^{-if_k} 
$$
then by unitary transformation we get from (\ref{eq:comm}) and 
(\ref{eq:anticomm}) that
\begin{equation}
	\left[ \widetilde {\cal D}_{M, (k)}, \; \widetilde  Q^{(k)}\right] = 0
\label{eq:tildecomm}
\end{equation}
\begin{equation}
	\left\{ \widetilde {\cal D}_{M, (k)}, \;  \sigma(\nu)\right\} 
	= 2\widetilde  Q^{(k)}-\sigma(\nu) 
\label{eq:tildeanticomm}
\end{equation}
on smooth sections. Since $ \widetilde {\cal D}_{M, (k)}$ has compact
resolvent, each eigenspace is finite dimensional and by
elliptic regularity consists of smooth sections. It then
follows from (\ref{eq:tildecomm}) and (\ref{eq:tildeanticomm})
that there
 exists an eigenbasis of $\wt{\cal D}_{M, (k)}$ consisting
of eigenspinors of $\wt Q^{(k)}$.

\begin{prop}\label{prop:qev}
The spectrum of $Q$ belongs to the set $\Z +c$.
Moreover, 
for any integer $k$, if 
\be 
	\wt Q^{(k)}\chi = (k+c)\chi,
\label{eq:Qeigenvalue}
\ee
then  there exists a section $\xi$
of the spinor bundle $\Psi_{m-k}$ on $N$ with Chern number $m-k$ such that
$\chi= \xi\circ\phi$.
\end{prop}

\begin{proof} Since $Q$ and $\wt Q^{(k)}$ are unitarily
equivalent for any $k$, it is enough to compute the spectrum of
$\wt Q^{(k)}$. Let $\wt Q^{(k)}\chi = E\chi$.
First we compute $\wt Q^{(k)}$ on any pull-up spinor $\eta\circ\phi$
with $\eta \in\Gamma(\Psi_{m-k})$. 
Notice that
$\wt Q^{(k)} = -i\widetilde\nabla^{M, (k)}_n
- \frac{1}{2}\sigma(\nu)$, hence
\begin{eqnarray}
	\wt Q^{(k)} (\eta\circ\phi)
	& = & -i \nabla_n^{M,(k)}(\eta\circ\phi)   +  (k+c)(\eta\circ\phi)
	 - \frac{1}{2}\sigma(\nu)(\eta\circ\phi) \nonumber \\
	& = &   (k+c)(\eta\circ\phi) 
\label{eq:Qoneta}
\end{eqnarray}
using (\ref{def:nablamn}), (\ref{def:tildenablam})
 and that $\phi^*(\nabla^{N, (k)})_n
(\eta\circ\phi)  =0$.

Next we choose a point $p\in \S^3$ where $\chi$ does not vanish.
We choose an orthonormal basis $\{ \xi^+, \xi^-\}$
 in $\Psi_{m-k}$ in a neighborhood $V$ around the point $\phi(p)$
and we pull it up. This gives an orthonormal basis  $\{ \xi^+\circ\phi,
 \xi^-\circ\phi\}$ in $\phi^{-1}(V)$, which is
a tubular neighborhood of the circle fiber $C$ going through $p$.
In this neighborhood
we can write the eigenspinor $ \chi$ as
$\chi = r_+ ( \xi^+\circ\phi) + r_-  ( \xi^-\circ\phi)$
with some functions $r_\pm\in C^\infty(M)$. Then
$$
	\wt Q^{(k)} \chi = - i (nr_+) ( \xi^+\circ\phi)
	- i (nr_-)  ( \xi^-\circ\phi) + (k+c)\chi
$$
and by $\wt Q^{(k)}\chi = E\chi$ and linear independence of
$\xi_\pm\circ\phi$ we get that $nr_\pm = i(E - (k+c))r_\pm$.
Hence $r_\pm$ must be of the form
$$
	r_\pm = r^{(0)}_\pm \exp \left[ i\theta\left(E - (k+c)
	\right)\right],
$$
where $\theta$ is the arclength parameter along $C$ in the direction
of $n$ and $n(r_\pm^{(0)})=0$.
 Since the total length of $C$ is $2\pi$ and at least one of
$r_+$, $r_-$ is not identically zero, we see that
$E-(k+c) \in \Z$, hence $E \in \Z +c$.

Now the second statement in  Proposition \ref{prop:qev} is
straight forward. If $E = k+c$  is
 an eigenvalue in (\ref{eq:Qeigenvalue}), then
$nr_+=nr_-=0$, i.e. $r_\pm$ are pull-up functions, $r_\pm = r^*_\pm\circ\phi$
with some $r^*_\pm\in C^\infty(N)$, hence $\chi = (r^*_+\xi_+)\circ\phi
+ (r^*_-\xi_-)\circ\phi$, and it is the pull-up of 
$\xi= r^*_+\xi_++r^*_-\xi_-$.
\end{proof}

We summarize our result

\begin{thm}\label{thm:spinorslift}
Let  ${\cal D}_M \psi = e\psi$ and $Q \psi = \mu \psi$.
Then $\mu = k+c$ with some $k\in \Z$. Fix this $k$, let
 $\wt\psi: = e^{if_k}\psi$, then by unitarity
\begin{equation}
	\wt {\cal D}_{M, (k)}
 \wt\psi = e\wt\psi \qquad\mbox{ and}\qquad
 \wt Q^{(k)} \wt\psi = \mu \wt\psi.
\label{eq:eveq}
\end{equation}
Then $\wt\psi= \xi\circ\phi$, with some section
 $\xi$
of the spinor bundle $\Psi_{m-k}$ on $N$ with Chern number $m-k$.
%and
%\begin{equation}
%	\wt{\cal D}_{M, (k)} \wt\psi 
%	= \wt{\cal D}_{M, (k)} \big(\xi\circ\phi\big)
%\label{eq:tildepullup}
%\end{equation}
Moreover, for any section $\chi$ of $\Psi_{m-k}$ we
have
\be
	\wt{\cal D}_{M, (k)} \big(\chi\circ\phi\big)
	= \left({\cal D}_{N, (k)} \chi\right)\circ\phi
	- \frac{1}{2} \chi\circ\phi + (k+c) \sigma(\nu)\chi\circ\phi\; .
\label{eq:diraconpullup}
\ee
\end{thm}

\begin{proof}
All statements have been proven in Proposition \ref{prop:qev}
except the (\ref{eq:diraconpullup}), which
is a straight forward calculation. 
\end{proof}

\subsection{Proof of Theorem \ref{thm:exact}}

\begin{proof} Let $e$ be an eigenvalue of ${\cal D}_M$ and consider
the corresponding eigenspace, which is finite dimensional.
In this subspace we find a simultaneous eigenbasis of $Q$, hence
we consider spinors $\psi$ with ${\cal D}_M\psi= e\psi$
and $Q\psi= \mu\psi$ for some $\mu$.  Then $\mu=k+c$
with some $k\in \Z$ and fix this $k$. Following 
Theorem~\ref{thm:spinorslift}, let $\wt\psi:= e^{if_k}\psi$
 and $\wt\psi = \xi\circ\phi$ with some $\xi\in \Psi_{m-k}$.
 {F}rom 
(\ref{eq:diraconpullup}) we have
\be
	\Big( \wt{\cal D}_{M, (k)} 
	+{1\over 2}\Big)\big(\xi\circ\phi\big)
	= \Big[\left({\cal D}_{N, (k)} + (k+c) \sigma(\nu)\right)\xi\Big]
	\circ\phi \; .
\label{eq:dmdn}
\ee
Using (\ref{eq:diraconpullup}) once more for
$\chi = \left({\cal D}_{N, (k)} + (k+c) \sigma(\nu)\right)\xi$ we obtain
$$
	\Big( \wt{\cal D}_{M, (k)} 
	+{1\over 2}\Big)^2\big(\xi\circ\phi\big)
	= \Big[\left({\cal D}_{N, (k)} + (k+c) \sigma(\nu)\right)^2\xi\Big]
	\circ\phi \; .
$$
Notice that $\{ {\cal D}_{N, (k)},  \sigma(\nu)\}=0$, which
easily follows from (\ref{eq:anticomm}) and (\ref{eq:Qoneta}).
In particular, the nonzero spectrum of ${\cal D}_{N, (k)}$ is
symmetric and
$$
	{\cal D}_{N, (k)}^2 \xi = \Big[ \Big( e+ {1\over 2}\Big)^2
	- (k+c)^2\Big] \xi
$$
i.e. $\xi$ is an eigenspinor of ${\cal D}_{N, (k)}^2$ and we define
$$
	\lambda:=
	\sqrt{\Big( e+ {1\over 2}\Big)^2- (k+c)^2}.
$$
If $\lambda >0$, 
then clearly $\lambda\in  \Sigma_+(k)$
(recall that $\Sigma_+(k)$ is the positive spectrum of ${\cal D}_{N, (k)}$).
The multiplicity of $e$ in the subspace $\{ \psi \; : \; Q\psi = (k+c)\psi\}$
 is bounded by the multiplicity of $\lambda$ in
the set $ \Sigma_+(k)$.

If $\lambda=0$, then by Theorem \ref{thm:AC}
the eigenvalue 0 belongs to the spectrum of
${\cal D}_{N, (k)}$ if and only if $m\neq k$. In this case
the multiplicity of the 0-eigenvalue is $|m-k|$, and the
eigenspinor is contained in the subspace $\{ \psi \; : \;\sigma(\nu)\psi = 
[\mbox{sgn}(m-k)]\psi\}$.
Hence, by (\ref{eq:dmdn}) we obtain $e = [\mbox{sgn}(m-k)] (k+c) -
\sfrac{1}{2}$ and the multiplicity of this eigenvalue
of $\wt{\cal D}_{M, (k)}$ is at most $|m-k|$.
This shows that $\mbox{Spec} \; {\cal D}_M$ is included in
the union given in (\ref{eq:dmspec}) with multiplicity.

For the converse statement, for any fixed 
$k\in \Z$ we start with an eigenspace
of ${\cal D}_{N, (k)}$ with eigenvalue $\lambda$.

If $\lambda =0$, then the same space is also an eigenspace of $\sigma(\nu)$
with eigenvalue $\mbox{sgn}(m-k)$ by Theorem \ref{thm:AC}.
Hence by (\ref{eq:dmdn}) the lift of this eigenspace to $\Psi_M$
is an eigenspace of $\wt{\cal D}_{M, (k)}$ with eigenvalue
$e= [\mbox{sgn}(m-k)](k+c) - \sfrac{1}{2}$.

If $\lambda >0$,  then for any element $\xi$  of this eigenspace
we form 
$$
	\chi^\pm : = \xi +  { -\lambda \pm \sqrt{\lambda^2+ (k+c)^2}
	\over k+c} \sigma(\nu)\xi
$$
if $k+c\ne 0$ and $\chi^+:=\xi$, $\chi^-:=\sigma(\nu)\xi$
if $k+c=0$. 
The sets $\{ \chi_j^- \}$
and $\{ \chi^+_j \}$ are both linearly independent 
as $\xi$'s run through a linearly independent set $\{ \xi_j\} \subset
\mbox{Ker }  \big( {\cal D}_{N, (k)} -\lambda \big)$.
For, if $\sum_j c_j \chi^+_j = 0$ with some constants $c_j$ then
$$
	0= \Big( 1 + \lambda^{-1}{\cal D}_{N, (k)}\Big)
	 \sum_j c_j \chi^+_j
	= 2 \sum_j c_j \xi^j .
$$

It is easy to check from (\ref{eq:dmdn})
that $\chi^\pm\circ\phi$ is an eigenspinor
of $\wt{\cal D}_{M, (k)}$ with eigenvalue $e= \pm\sqrt{\lambda^2
+(k+c)^2} - \sfrac{1}{2}$. This completes the proof of
Theorem \ref{thm:exact}.

\end{proof}

\appendix
\section{Appendix: 
Spinor bundles on $\S^2$ (and $\S^3$) and the Aharonov-Casher Theorem}

We shall now construct spinor bundles $\Psi$ on $\S^2$.
First we choose coordinates. Let $\S^2_+=\S^2\setminus\{S\}$ and 
$\S^2_-=\S^2\setminus\{N\}$, where $N$ and $S$ are the north and 
south poles respectively. 

Consider the stereographic projections
$z_\pm:\S^2_\pm\to\C$ defined by
$$
  \omega=\left\{
    \begin{array}{ll}\displaystyle
      \left(\frac{-4\overline{z_-(\omega)}}{4+|z_-(\omega)|^2},\
    -\frac{4-|z_-(\omega)|^2}{4+|z_-(\omega)|^2}\right),
  &\hbox{for }\omega\in\S^2_-\\ \\ \displaystyle
\left(\frac{4z_+(\omega)}{4+|z_+(\omega)|^2},\
    \frac{4-|z_+(\omega)|^2}{4+|z_+(\omega)|^2}\right),
  &\hbox{for }\omega\in\S^2_+
\end{array}\right.
,
$$ 
where we have identified $\S^2\subset \C\times \R$.
Note that for $\omega\in\S^2_-\cap\S^2_+$ we have
$z_-(\omega)=-4{z_+(\omega)}^{-1}$. 
With the above choice the maps $z_\pm$ are 
orientation preserving when we choose the standard orientations of
$\S^2$ and $\C$ (strictly speaking $z_-$ is a stereographic
projection followed by a reflection).
If we use the metric 
$$
    ds^2= (1+\sfrac{1}{4}|z|^2)^{-2}dz\, d\overline{z}
$$
on $\C$ both maps $z_\pm$ are also isometries.

\subsection{Spinor bundles on $\S^2$ and $\S^3$}\label{sec:spinons2}
Corresponding to each $n\in\Z$ we define a spinor bundle
$\Psi_n$ on $\S^2$ by the following properties:
\begin{itemize}
\item There are open subsets $\Psi_n^\pm\subset\Psi_n$ such  that 
$\Psi_n=\Psi_n^+\cup\Psi_n^-$
\item There are diffeomorphisms
$\phi_\pm:\Psi^{(\pm)}_n\to\S^2_\pm\times\C^2$.
\item If $\eta\in\Psi_n$ and $\phi_\pm(\eta)=(\omega_\pm,u_\pm)$ then
\begin{equation}\label{eq:psin}
        \omega_+=\omega_-\quad\hbox{and}\quad
        u_-={\cal U}_n(z_+(\omega_+))
        {\cal W}(z_+(\omega_+))u_+
\end{equation}
where
$$
{\cal U}_n(z)=\left(\frac{|z|}{z}\right)^n
\quad\mbox{and}\quad
     {\cal W}(z)=
     \left(\begin{array}{cc}
       z|z|^{-1}&0\\ 
       0&\overline{z}|z|^{-1}
     \end{array}\right)\in SU(2)
$$
\item If $\alpha=a(z)d\overline{z}+\overline{a(z)}dz$ 
is a real one-form on $\C$ then the Clifford multiplication $\sigma$
on $\Psi_n$ is defined by 
$$
        \phi_\pm\left(\sigma\left(z^*_\pm(\alpha)\right)\eta\right)
        =\left(\omega,\ \left(1+\sfrac{1}{4}|z_\pm(\omega)|^2\right)
        \left(\begin{array}{cc}
       0&\overline{a(z_\pm(\omega))}\\ 
       a(z_\pm(\omega))&0
     \end{array}\right)u_\pm\right),
$$
when $\phi_\pm(\eta)=(\omega,u_\pm)$. Here $z^*_\pm(\alpha)$
are the pull-backs of the one-form $\alpha$ to $\S^2_\pm$.
Note that it is the Clifford multiplication relative to 
the metric $ds^2$ which is being used on $\C^2$.
\end{itemize}

It is fairly easy  to check that this really defines a spinor bundle 
$\Psi_n$ on $\S^2$. 
In particular, we notice that if $\tilde{\alpha}$ is a one-form
on $\S^2$ then $\tilde{\alpha}_{\Bigl|\S^2_\pm}=z^*_\pm(\alpha_\pm)$, 
where $\alpha_\pm=a_\pm(z)\overline{dz}+\overline{a_\pm(z)}dz$ 
satisfies
$a_+(z)=4a_-(-4z^{-1})\overline{z}^{-2}$. Thus we see that 
the Clifford multiplication transforms consistently 
between $\Psi^{(\pm)}$, i.e., 
\begin{eqnarray*}
\lefteqn{(1+\sfrac{1}{4}|z_-|^2)\left(\begin{array}{cc}
       0&\overline{a_-(z_-)}\\ 
       a_-(z_-)&0
     \end{array}\right)}\hspace{3truecm}&&\\ \\
&=&(1+\sfrac{1}{4}|z_+|^2){\cal W}(z_+)\left(\begin{array}{cc}
       0&\overline{a_+(z_+)}\\ 
       a_+(z_+)&0
     \end{array}\right){\cal W}(z_+)^*
\end{eqnarray*}
where $z_\pm=z_\pm(\omega)$.

\begin{prop}\label{prop:isom} If $\Psi$ is a spinor bundle over $\S^2$ 
then $\Psi$ is diffeomorphic to $\Psi_n$ for some $n\in\Z$.
\end{prop}
\begin{proof}
We just sketch this standard argument. 
Since any vector bundle on $\C$ or on $\S^2_\pm$ 
is trivial we easily see that 
any spinor bundle on $\S^2$ is of 
the form described above with ${\cal U}_n(z_+(\omega))$ replaced 
by a general function ${\cal U}:\S^2\setminus\{N,S\}\to U(1)$. 
Let $-n$ be the degree of the map 
${\cal U}$ (i.e., the degree when we restrict to e.g.\ the equatorial circle). 
Then $\S^2\setminus\{N,S\}\ni \omega\mapsto 
{\cal U}(\omega){\cal U}_n^*(z_+(\omega))\in U(1)$
is a map of degree 0. We may therefore find two functions
${\cal U}_\pm:\S^2\setminus\{N,S\}\to U(1)$ with
${\cal U}_+$  equal 1 near $N$ and likewise for $S$ such that 
$$
  {\cal U}(\omega){\cal U}_n^*(z_+(\omega))={\cal U}_+(\omega)
                {\cal U}_-(\omega)^*.
$$
If we now use ${\cal U}_\pm$ to change the coordinates
on the fibers of $\Psi$ over $\S^2_\pm$ respectively
we see that the transformation matrix ${\cal U}$ will be replaced 
by ${\cal U}_n\circ z_+$.
\end{proof}

\begin{remark}
A similar argument immediately proves that on $\S^3$ there is only
one $Spin^c$ bundle. This bundle is in fact trivial, since
we can find a global orthonormal frame $(e_1, e_2, e_3):=
(u_1, u_2, n)$ on $\S^3$
(see Section \ref{sec:hopf}). The Clifford multiplication
on $\S^3\times \C^2$ is defined by $\sigma(e^j):=\sigma_j$, where
$\sigma_j$ is the $j$-th Pauli matrix.
\end{remark}

\subsection{$Spin^c$ connections on $\Psi_n$}\label{sec:spincons2}

We first describe $Spin^c$ connections on $\C$ with respect to the 
metric $ds^2=(1+\frac{1}{4}|z|^2)^{-2}dzd\overline{z}$. 
On $\C$ we consider the trivial spinor bundle, i.e., 
$\Psi=\C\times\C^2$. 

Since the metric $ds^2$  is conformally equivalent to the 
standard metric $dzd\overline{z}$ with the conformal factor 
$\Omega(z)=(1+\frac{1}{4}|z|^2)^{-1}$ we may use the results 
from Section~\ref{sec:conformal}.
The spinor sections with respect to the standard metric on 
$\C$ are given in terms of a (real) one-form 
$\alpha=a(z)d\overline{z}+\overline{a(z)}dz$ as follows. 
The covariant derivative  
along a (real) vector field
$X=\xi(z)\partial_z+\overline{\xi(z)}\partial_{\overline{z}}$ 
is
\begin{equation}\label{eq:standardnabla}
   \xi(z)\Bigl(\partial_z-i\overline{a(z)}\Bigr)
   +\overline{\xi(z)}\Bigl(\partial_{\overline{z}}-i{a(z)}\Bigr).
\end{equation}
We may calculate the covariant derivative of spinors 
in the conformal 
metric $ds^2$ from
Prop~\ref{prop:conformalspinconnection}
using
$$
   \sigma(X^*)=\sfrac{1}{2}
   \sigma\left(\overline{\xi(z)}dz+{\xi(z)}d\overline{z}\right)
   =\left(\begin{array}{cc}0&\overline{\xi(z)}\\ \xi(z)&0
         \end{array}\right)
 $$ 
and 
$$
  \sigma(d\Omega)=-\sfrac{1}{2}
  (1+\sfrac{1}{4}|z|^2)^{-2}\left(\begin{array}{cc}0&\overline{z}\\ 
      z&0
         \end{array}\right).
$$
According to Prop~\ref{prop:conformalspinconnection} the
covariant derivative of spinors in the 
metric $ds^2$ is
$$
  \nabla^{\alpha}_X=\xi(z)\Bigl(\partial_z-i\overline{a(z)}\Bigr)
   +\overline{\xi(z)}\Bigl(\partial_{\overline{z}}-i{a(z)}\Bigr)
   -\frac{i\mbox{Im}\left(z\overline{\xi(z)}\right)}{(4+|z|^2)}
   \left(\begin{array}{cc}1&0\\ 
      0&-1
         \end{array}\right).
$$

Now assume that we have a covariant derivative $\nabla$ on 
$\Psi_n$. Let $\eta$ be a section in $\Psi_n$ and $\tilde X$ 
a vectorfield on $\S^2$. For $\omega\in\S^2_\pm$ write 
$$
    \phi_\pm(\eta(\omega))=(\omega,u_\pm(z_\pm(\omega)))
    \quad\mbox{and}\quad (z_\pm)_*\tilde
    X(\omega)=(\omega,X_\pm(z_\pm(\omega))).
$$ 
Then there exists one-forms $\alpha_\pm$ on $\C$ such that 
\begin{equation}\label{eq:nablaS2def}
    \phi_\pm\left(\nabla_{\tilde X}\eta(\omega)\right)=\left(\omega,
    \nabla^{\alpha_\pm}_{X_\pm} u_\pm(z_\pm(\omega)) \right).
\end{equation}

Note that by (\ref{eq:psin}) we must have
$$
   \nabla^{\alpha_-}_{X_-} u_-(z_-(\omega)) = 
   \left(\frac{|z_+(\omega)|}{z_+(\omega)}\right)^n 
        {\cal W}(z_+(\omega))\nabla^{\alpha_+}_{X_+}u_+(z_+(\omega))
$$
i.e.,
\begin{equation}\label{eq:nablatransform}
   \nabla^{\alpha_-}_{X_-}\left[\left({\cal U}_n
        {\cal W}u_+\right)(-4z^{-1})\right]= 
  {\cal U}_n(-4z^{-1})
        {\cal W}(-4z^{-1})\left(\nabla^{\alpha_+}_{X_+}u_+\right)(-4z^{-1})
\end{equation}
With $X_\pm=2\mbox{Re}\left[\xi_\pm(z)\partial_z\right]$
we have the relation
$
  \xi_+(z)=\sfrac{1}{4}z^2\xi_-(-4z^{-1}).
$
A straightforward calculation then shows that
if $\alpha_\pm=2\mbox{Re}\left[a_\pm(z)d\overline{z}\right]$
then (\ref{eq:nablatransform}) implies
that 
\begin{equation}\label{eq:apmrelation}
  a_-(z)=4\overline{z}^{-2}a_+(-4z^{-1})+i\sfrac{n}{2}\overline{z}^{-1}.
\end{equation}
Conversely, for any choice of functions $a_\pm$ on $\C$ satisfying 
(\ref{eq:apmrelation}) the relation (\ref{eq:nablaS2def})
will define a $Spin^c$ connection on $\S^2$.

It is easy to see that
$\beta_{\bigl|\S^2_\pm}=z_\pm^*(d\alpha_\pm)$.
Using Stokes law we then have that
\be
          \int_{\S^2}\beta=\int_{C}z_-^*(\alpha_-)+
        z_+^*(\alpha_+)=\frac{in}{2}\int_{|z|=2}
        \frac{d\overline{z}}{\overline{z}}-
        \frac{d{z}}{{z}}=2\pi n \; ,
\label{eq:fieldn}
\ee
where $C$ is the equatiorial curve oriented appropriately, 
corresponding to the circle $|z|=2$ being oriented 
counterclockwise. Note that 
$$
	\int_{|z|=2} 4\overline{z}^{-2} a_+(-4z^{-1}) 
	d\overline{z} = -\int_{|z|=2} a_+(z)d\overline{z}.
$$

\begin{prop}\label{prop:anyfield}
For any closed 2-form $\beta$ on $\S^2$ with $\int\beta=2\pi n$
there is a $Spin^c$ connection on $\Psi_n$ such that 
$\beta$ is the magnetic 2-form. 
\end{prop}
\begin{proof}
We must show that there are one-forms $\alpha_\pm$ on $\C$ 
satisfying (\ref{eq:apmrelation})
such that $\beta_{\bigl|\S^2_\pm}=z_\pm^*(d\alpha_\pm)$.
We construct them explicitly.
As before we will write $\alpha_\pm=2\mbox{Re}[a_\pm(z)d\overline{z}]$
with $a_\pm$ defined below.

Let $\tilde\beta_\pm=(z_\pm^{-1})^*(\beta)$.
We define
\begin{equation}\label{eq:h+}
	h_+(z) := \pi^{-1}\int_{\C} \log |z-z'|^2 \tilde\beta_+ (z') \; ,
\end{equation}
$$
	a_+(z) : = {i\over 4}  \partial_{\overline{z}} h_+(z) 
$$
and
$$
	a_-(z):=
	4\overline{z}^{-2}a_+(-4z^{-1})+i\sfrac{n}{2}\overline{z}^{-1}
$$
according to (\ref{eq:apmrelation}).
A simple calculation shows that $a_-(z)$ is smooth on $\C$, i.e.
the singularities apparently present in (\ref{eq:apmrelation})
exactly cancel each other. For this argument we use that $\int_\C \beta_+
=2\pi n$ and that $\beta_+$ is the pushforward of a (smooth)
2 form on $\S^2$.

{F}rom the definition of $h_+$ we see that 
 $\big( \partial_{\overline{z}}\partial_z h(z)\big) dz\wedge
d\overline{z}
=-2i\tilde\beta_+(z)$, which implies
  $d\alpha_+=\tilde\beta_+$. Finally $d\alpha_-=\tilde\beta_-$
follows from the relation (\ref{eq:apmrelation}) and that
$z^*_-(d\alpha_-) = z^*_+(d\alpha_+)$ on $\S^2_-\cap \S^2_+$.

\end{proof}

\subsection{The Dirac operator on $\Psi_n$}
Let ${\cal D}$ be the Dirac operator corresponding
to the $Spin^c$ connection $\nabla$ on a
spinor bundle $\Psi_n$ on $\S^2$ with Chern number $n$.
 We may then for any spinor field
$\eta$ in $\Psi_n$ write
$$
        \phi_\pm(\tilde{\cal D}\eta(\omega))
        =\left(\omega, {\cal D}_\pm u_\pm(z_\pm(\omega))\right)
$$
for $\omega\in\S^2_\pm$. Here ${\cal D}_\pm$ are the Dirac
operators on $\C$ corresponding to the metric $ds^2$. 
According to Theorem~\ref{thm:conformaldirac}
we can express ${\cal D}_\pm$
in terms of the standard Dirac operators on $\C$
corresponding to the standard covariant derivative
(\ref{eq:standardnabla}). The standard Dirac operators are 
\begin{equation}\label{eq:standarddirac}
          -2i\left(
        \begin{array}{cc}0&\partial_z-i\overline{a_\pm(z)}\\    
        \partial_{\overline{z}}-i{a_\pm(z)}&0
        \end{array}\right).
\end{equation}

Finally, we give the
proof of the Aharonov Casher Theorem stated in Section \ref{sec:exact}.

{\it Proof of Theorem \ref{thm:AC}.}
 Let $\eta\in\mbox{ker}{\cal D}$ and  define
$u_\pm:\C\to\C^2$ by  
$\phi_\pm\eta(\omega)=(\omega,u_\pm(z_\pm(\omega)))$, for 
$\omega\in\S^2_\pm$. Then ${\cal D}_\pm u_\pm=0$.
According to Theorem~\ref{thm:conformaldirac} we then
have that 
$
         v_\pm(z)=\Omega(z)^{1/2}u_\pm(z),
$ 
where as before $\Omega(z)=(1+\frac{1}{4}|z|^2)^{-1}$,
are in the kernel of the standard Dirac operators 
(\ref{eq:standarddirac}). 
By 
(\ref{eq:psin}) the spinors $v_\pm$ 
satisfy the transformation property 
\begin{equation}\label{eq:trans}
 \Omega(z)^{-1/2}v_-(z)=\Omega(-4z^{-1})^{-1/2}
{\cal U}(-4z^{-1}){\cal W}(-4z^{-1})v_+(-4z^{-1}) \; .
\end{equation}

Clearly the map from $\eta$ to $(v_-,v_+)$ with the above 
transformation property is 
a linear isomorphism. 

It turns out to be fairly simple to characterize
the elements $v_\pm$ in the kernel 
of the standard Dirac operators. Since $ \sigma_3$
anticommutes with the standard Dirac operator, we
 can consider the cases
$\sigma_3 v_\pm = v_\pm$ and $\sigma_3 v_\pm = -v_\pm$ separately.
Assume for definiteness that $\sigma_3 v_\pm = v_\pm$, i.e.
\begin{equation}\label{eq:standard}
[\partial_{\bar z} - ia_\pm(z)] v_\pm =0 \; .
\end{equation}
 
We write $v_\pm$ in the form
$$
    	v_\pm(z) = f_\pm(z) e^{-\sfrac{1}{4} h_\pm(z)}
$$
where $h_+$ is defined in (\ref{eq:h+}) and
$$
	h_-(z) : = h_+ (-4z^{-1}) + 2n \log |z|^2 \; .
$$
One easily computes that $\partial_{\bar z}  h_\pm(z) = ia_\pm(z)$,
and using (\ref{eq:standard}) we obtain that $\partial_{\bar z} f_\pm (z)=0$,
i.e. these are analytic functions. Finally (\ref{eq:trans})
gives the following relation between $f_-$ and $f_+$
$$
	f_-(z) = 2(-z)^{n-1} f_+(-4z^{-1}) \; .
$$
Hence $f_-$ is an analytic function, bounded by a constant
times $|z|^{n-1}$ at infinity. Then $n\ge 1$ and $f_-$ is a polynomial
of degree at most $n-1$. A basis in the kernel of the Dirac
operator is obtained by choosing $f_-(z) = 1, z, \ldots z^{n-1}$,
and the dimension is $n$. 

Similar argument shows that if $\sigma_3 v_\pm = -v_\pm$, then
 $n\leq -1$, and the dimension of the space of such zero modes
is $|n|$. In particular all zero modes have definite spin
and only one of the two eigenspaces of $\sigma_3=\sigma(\nu)$
can accomodate zero modes, depending on the sign of $n$.
Recalling that $n$ is the total flux divided by $2\pi$,
we have completed the proof of Theorem \ref{thm:AC}.
\qed

\noindent
{\it Address of the authors:}

\noindent
{\it L\'aszl\'o Erd\H os}: School of Mathematics, Georgia Institute of
Technology, Atlanta GA 30332, USA.

\noindent
{\it E-mail}: lerdos@math.gatech.edu

\bigskip
\noindent
{\it Jan Philip Solovej}: Department of Mathematics, University
of Copenhagen, Universitetsparken 5, DK-2100 Copenhagen, Denmark.

\noindent
{\it E-mail}: solovej@math.ku.dk

\end{document}